%% file: ms.tex
\documentclass{emulateapj}

\slugcomment{SUBMITTED to ApJ, 21 February 2006; ACCEPTED, 8 May 2006}
\shorttitle{Dim Suns 1}
\shortauthors{Cushing et al.}

\begin{document}


\title{A \textit{Spitzer} Infrared Spectrograph (IRS) Spectral Sequence of M, L, and T Dwarfs}


\author{Michael C. Cushing\altaffilmark{1,2},
        Thomas L. Roellig\altaffilmark{3},
        Mark S. Marley\altaffilmark{4},
        D. Saumon\altaffilmark{5},
        S.~K. Leggett\altaffilmark{6},
        J. Davy Kirkpatrick\altaffilmark{7},
        John C. Wilson\altaffilmark{8},
        G.~C. Sloan\altaffilmark{9},
        Amy K. Mainzer\altaffilmark{10}, 
        Jeff E. Van Cleve\altaffilmark{11}, \&
        James R. Houck\altaffilmark{12}}

\altaffiltext{1}{Steward Observatory, University of Arizona, 933 North Cherry Avenue, Tucson, AZ 85721, mcushing@as.arizona.edu}

\altaffiltext{2}{Spitzer Fellow}

\altaffiltext{3}{NASA Ames Research Center, MS 245-6, Moffett Field, CA 94035, thomas.l.roellig@nasa.gov}

\altaffiltext{4}{NASA Ames Research Center, MS 254-3, Moffett Field, CA 94035, mmarley@mail.arc.nasa.gov}

\altaffiltext{5}{Los Alamos National Laboratory, Applied Physics Division, MS P365, Los Alamos, NM 87544, dsaumon@lanl.gov}

\altaffiltext{6}{Joint Astronomy Centre, University Park, Hilo, HI 96720, s.leggett@jach.hawaii.edu}

\altaffiltext{7}{Infrared Processing and Analysis Center, MC 100-22, California Institute of Technology, Pasadena, CA 91125, davy@ipac.caltech.edu}

\altaffiltext{8}{Astronomy Building, University of Virginia, 530 McCormick Road, Charlottesville, VA 22903, jcw6z@virginia.edu}

\altaffiltext{9}{Astronomy Department, Cornell University, Ithaca, NY 14853, sloan@isc.astro.cornell.edu}

\altaffiltext{10}{Jet Propulsion Laboratory, MS 169-506, 4800 Oak Grove Drive, Pasadena, CA 91109, amainzer@jpl.nasa.gov}

\altaffiltext{11}{Ball Aerospace and Technologies Corporation, 1600 Commerce Street, Boulder, CO 80301, jvanclev@ball.com}

\altaffiltext{12}{Astronomy Department, Cornell University, Ithaca, NY 14853, jrh13@cornell.edu}

\begin{abstract}

  We present a low-resolution ($R$ $\equiv$ $\lambda/\Delta\lambda$
  $\approx$ 90), 5.5$-$38 $\mu$m spectral sequence of a sample of M,
  L, and T dwarfs obtained with the Infrared Spectrograph (IRS)
  onboard the \textit{Spitzer Space Telescope}.  The spectra exhibit
  prominent absorption bands of H$_2$O at 6.27 $\mu$m, CH$_4$ at 7.65
  $\mu$m, and NH$_3$ at 10.5 $\mu$m and are relatively featureless at
  $\lambda \gtrsim$ 15 $\mu$m.  Three spectral indices that measure
  the strengths of these bands are presented; H$_2$O absorption
  features are present throughout the MLT sequence while the CH$_4$
  and NH$_3$ bands first appear at roughly the L/T transition.
  Although the spectra are, in general, qualitatively well matched by
  synthetic spectra that include the formation of spatially
  homogeneous silicate and iron condensate clouds, the spectra of the
  mid-type L dwarfs show an unexpected flattening from roughly 9 to 11
  $\mu$m.  We hypothesize that this may be a result of a population of
  small silicate grains that are not predicted in the cloud models.
  The spectrum of the peculiar T6 dwarf 2MASS J0937$+$2931 is
  suppressed from 5.5$-$7.5 $\mu$m relative to typical T6 dwarfs and
  may be a consequence of its mildly metal-poor/high surface gravity
  atmosphere.  Finally, we compute bolometric luminosities of a
  subsample of the M, L, and T dwarfs by combining the IRS spectra
  with previously published 0.6$-$4.1 $\mu$m spectra and find good
  agreement with the values of Golimowski et al. who use $L'$- and
  $M'$-band photometry and to account for the flux emitted at $\lambda
  >$ 2.5 $\mu$m.

\end{abstract}


\section{Introduction}

The discovery of the first bona fide brown dwarf (BD) Gl 229B
\citep{1995Natur.378..463N} and the confirmation of other BD candidates
\citep{1996ApJ...458..600B,1996ApJ...469L..53R} ushered in a new era in
both stellar and planetary astrophysics since BDs bridge the gap in mass
between stars and planets.  Over four hundred very low-mass stars and
BDs, collectively known as ``ultracool'' dwarfs, have since been
discovered, primarily in wide-field optical and near-infrared surveys
such as the Two Micron All Sky Survey
\citep[2MASS;][]{2006AJ....131.1163S}, the Deep Near Infrared Southern
Sky Survey \citep[DENIS;][]{1997Msngr..87...27E}, and the Sloan Digital
Sky Survey \citep[SDSS;][]{2000AJ....120.1579Y}.  Ultracool dwarfs have
effective temperatures ($T_{\mathrm{eff}}$) less than $\sim$2700 K and
spectral types later than $\sim$M7 V, and include the new L and T
dwarfs.  Since their spectral energy distributions (SEDs) peak near
$\sim$1 $\mu$m, considerable observational \citep[][Kirkpatrick
2005]{2000ARA&A..38..485B} and theoretical
\citep{2000ARA&A..38..337C,2001RvMP...73..719B} effort has gone into
studying them at both red-optical and near-infrared wavelengths.

Nevertheless, observations at $\lambda >$ 2.5 $\mu$m can also provide
important constraints on the fundamental parameters and atmospheric
physics of ultracool dwarfs.  For example, the $T_{\mathrm{eff}}$ of an
ultracool dwarf is typically determined by combining an observational
bolometric luminosity with a theoretical radius
\citep[e.g.,][]{2002AJ....124.1170D,2004AJ....127.3516G}.  Since
spectroscopic observations are typically limited to $\lambda \lesssim$
2.5 $\mu$m, the $T_{\mathrm{eff}}$ scale therefore depends critically on
an accurate accounting of the flux emitted at longer wavelengths.  The
effects of non-equilibrium chemistry on the abundances of CO, CH$_4$,
N$_2$, and NH$_3$ due to the vertical transport of gas within the
atmospheres of ultracool dwarfs, and hence the band strengths of CO,
CH$_4$, and NH$_3$, are also strongest at these wavelengths
\citep{saumon03,2003IAUS..211..345S}.  In addition, observations at
$\lambda >$ 2.5 $\mu$m are easier to interpret using atmospheric models
because the dominant absorption bands (H$_2$O, CO, CH$_4$, and NH$_3$)
at these wavelengths arise from fundamental transitions with nearly
complete line lists compared to the overtone and combination bands at
near-infrared wavelengths.  Finally, mid-infrared spectroscopy adds
important information about the vertical structure and properties of
atmospheric condensates.  In principle, Mie scattering effects of iron
and silicate grains expected in the atmospheres of the L dwarfs allow
constraints to be placed on particle sizes, but only with spectra
obtained over a large wavelength range.  In addition, if a population of
small particles is present, silicate absorption features may be apparent
near 10 $\mu$m.

Unfortunately, observations of ultracool dwarfs at $\lambda >$ 2.5
$\mu$m are not as common as they are at shorter wavelengths due to the
difficulty of observing from the ground at these wavelengths.  The
majority of the observations consist of $L'$ and $M'$-band photometry
\citep{2001ApJ...556L..97S,2002ApJ...564..452L,2004AJ....127.3516G},
although some spectroscopy has been performed at these wavelengths
\citep{1997ApJ...489L..87N,1998ApJ...502..932O,2001PhDT.......116B,2000ApJ...541L..75N,2005ApJ...623.1115C}.
In particular, the $\nu_3$ fundamental band of CH$_4$ at $\sim$3.3
$\mu$m has been detected in the spectra of L and T dwarfs
\citep{1998ApJ...502..932O,2001PhDT.......116B,2000ApJ...541L..75N,2005ApJ...623.1115C}
and the fundamental CO band at $\sim$4.7 $\mu$m has been detected in the
T dwarf Gl 229B \citep{1997ApJ...489L..87N,1998ApJ...502..932O}.  Even
fewer ground-based observations of ultracool dwarfs exist at $\lambda >$
5 $\mu$m and are limited to photometric observations
\citep{1996AJ....112.1678M,2004ApJ...602L.129C,2005A&A...436L..39S} of
just a few dwarfs.

The launch of the \textit{Spitzer Space Telescope}
\citep{2004ApJS..154....1W}, which is sensitive from 3.6 to 160
$\mu$m, has opened a heretofore untapped wavelength range for the
study of ultracool dwarfs.  In particular, the Infrared Array Camera
\citep[IRAC, ][]{2004ApJS..154...10F} and the Infrared Spectrograph
\citep[IRS, ][]{2004ApJS..154...18H} are providing unprecedented
photometric and spectroscopic observations of ultracool dwarfs at
mid-infrared wavelengths.  \citet{2004ApJS..154..418R} presented the
first mid-infrared spectra of an M, L, and T dwarf and identified
absorption bands of H$_2$O, CH$_4$, and NH$_3$.  In this paper we
extend the work of \citeauthor{2004ApJS..154..418R} and present a
5.3$-$38 $\mu$m spectral sequence of M, L and T dwarf spectra obtained
with the IRS.  Forthcoming papers will provide a more in-depth
analysis of the spectra.  We describe the observations, data
reduction, and absolute flux calibration of the spectra in \S2 while
in \S3 we discuss the spectra sequence, derive three spectral indices
that measure the strengths of the H$_2$O, CH$_4$, and NH$_3$ bands,
and discuss a number of interesting object.  In \S4 we present full
0.6$-$15 $\mu$m spectral energy distributions of a subsample of M, L,
and T dwarfs and compute their bolometric luminosities.

\section{Observations and Data Reduction}

Our current sample consists of 14 M dwarfs, 21 L dwarfs, and 11 T
dwarfs drawn from the literature\footnote{Databases of known L and T
  dwarfs can be found at \url{http://DwarfArchives.org} and
  \url{http://www.jach.hawaii.edu/~skl/LTdata.html}} .  The
observations were conducted with the IRS as part of the ``Dim Suns''
IRS Science Team Guaranteed Time Observation (GTO) program.  The IRS
is composed of four modules capable of performing low- ($R$ $\equiv$
$\lambda/\Delta\lambda$ $\approx$ 90) to moderate-resolution ($R$
$\approx$ 600) spectroscopy from 5.3 to 38 $\mu$m.  We used the
Short-Low (SL) module that covers from 5.3 to 15.3 $\mu$m at $R$
$\approx$ 90 in two orders and the Long-Low (LL) module that covers
from 14.0 to 38 $\mu$m at $R$ $\approx$ 90 also in two orders.  A log
of the observations, including the \textit{Spitzer} AOR key,
spectroscopic module, and total on-source integration time is given in
Table \ref{obslog}.  Although both optical and infrared spectral types
are listed in Table \ref{obslog}, we hereafter use optical types for
the M and L dwarfs \citep{1991ApJS...77..417K,1999ApJ...519..802K} and
infrared types for the T dwarfs \citep{2006ApJ...637.1067B} unless
otherwise noted.  In addition, we hereafter abbreviate the numerical
portions of the 2MASS, SDSS, and DENIS target designations as
Jhhmm$\pm$ddmm, where the suffix is the sexigesimal Right Ascension
(hours and minutes) and declination (degrees and arcminutes) at J2000
equinox.

The observations consisted of a series of exposures obtained at two
positions along each slit.  The raw IRS data were processed with the
IRS pipeline (version S12) at the \textit{Spitzer} Science Center.
The data were reduced using custom IDL procedures based on the
Spextool \citep{2004PASP..116..362C} data reduction package.  The
background signal was first removed from each science frame by
subtracting the median of the frames obtained in the same
spectroscopic module but with the target in the other order.  Any
residual background was removed by subtracting off the median signal
in the slit at each column, excluding regions that contain signal from
the target.  The spectra were then extracted with a fixed-width
aperture (6$\farcs$0 in the SL module and 9$\farcs$0 in the LL module)
and wavelength calibrated using the technique employed by the IRS data
reduction package, the Spectroscopic Modeling Analysis and Reduction
Tool \citep[SMART; ][]{2004PASP..116..975H}.

Observations of standard stars obtained as part of the normal IRS
calibration observations were used to remove the instrument response
function and flux calibrate the science targets.  We used $\alpha$ Lac
(A0 V) to correct the SL spectra and HR 6348 (K1 III) for the LL
spectra.  The standard star spectra were extracted in a similar
fashion to the science targets.  Model spectra of the two standard
stars from \citet{2003AJ....125.2645C} were used to remove the
intrinsic stellar energy distribution from the raw standard star
spectra.  The spectra from the SL module were then merged into a
single 5.3$-$15.3 $\mu$m spectrum and the spectra from the LL module
were merged into a single 14.0$-$38 $\mu$m spectrum.  Finally, for
those targets with both SL and LL data, the spectra from the two
modules are merged together; any offset in the flux density levels of
the two spectra are removed by scaling the LL spectrum to the flux
density level of the SL spectrum.

The final step in the reduction process is to absolutely flux
calibrate the spectra using IRAC Band 4 photometry.  Fortunately, 27
of the dwarfs in our sample have been observed as part of an IRAC
Science Team GTO program (B. Patten, in preparation).  Below we
describe the process used to absolutely flux calibrate the dwarf
spectra in our sample.

%\textit{assuming a dangling participle, and didn't define nu_0, and
%Burgasser astroph ref 2005a}

IRAC observations are reported as a flux density,
$f^{\mathrm{IRAC}}_{\nu}(\lambda_0)$, at a nominal wavelength
$\lambda_0$ ($\lambda_0$ = 7.872 $\mu$m for Band 4 \citep{2005PASP..117..978R}) \textit{assuming}
the target has a spectral energy distribution given by,

\begin{equation}
\nu \widetilde{f}_{\nu}(\nu) = \mathrm{constant} = \nu_0
f_{\nu}^{\mathrm{IRAC}}(\lambda_0), 
\end{equation}
\noindent
where $\nu_0 = c / \lambda_0$.  If this assumption is invalid, as it is
for ultracool dwarfs, then the quoted flux density is not the flux
density of the target at $\lambda_0$.  Therefore in order to compare an
IRAC observation to an IRS spectrum, we must compute an equivalent
$f^{\mathrm{IRAC}}_{\nu}(\lambda_0)$ given the IRS spectrum.

$f^{\mathrm{IRAC}}_{\nu}(\lambda_0)$ is determined for any source with
a SED given by $f_{\nu}(\lambda)$ from the requirement that the number
of electrons detected per second from the source, $N_e$, be equal to
the number of electrons detected per second from a hypothetical target
with a SED given by Equation 1, $\widetilde{N}_e$.  That is,

\begin{equation}
A \int \frac{\widetilde{f}_{\nu}(\nu)}{h\nu}S(\nu) d\nu = A \int \frac{f_{\nu}(\nu)}{h\nu}S(\nu) d\nu, 
\end{equation}
\noindent
where $A$ is the area of the telescope and $S(\nu)$ is the system
response function of the telescope plus instrument plus detector
system in units of photon electron$^{-1}$.  Substituting Equation 1
into Equation 2, and solving for $f^{\mathrm{IRAC}}_{\nu}(\lambda_0)$,
we find

\begin{equation}
f^{\mathrm{IRAC}}_{\nu}(\lambda_0) = \frac{\int (\nu_{0} / \nu) f_{\nu}(\nu) S(\nu
) d\nu}{\int ( \nu_{0} / \nu)^{2}S(\nu) d\nu}.
\end{equation}
\noindent
Equation 3 can be used to predict the flux density IRAC would report
if it were to observe a source with a SED given by $f_{\nu}(\nu)$.

The IRS spectra of the dwarfs with IRAC observations were absolutely
flux-calibrated by multiplying each spectrum by a scale factor $C$
given by,

\begin{equation}
C = f^{\mathrm{IRAC}}_{\nu}(\lambda_0) / f^{\mathrm{IRS}}_{\nu}(\lambda_0), 
\end{equation}
where $f^{\mathrm{IRS}}_{\nu}(\lambda_0)$ was determined using the IRS
spectrum and Equation 4 and $f^{\mathrm{IRAC}}_{\nu}(\lambda_0)$ is
the reported IRAC flux density for the dwarf in question.  The
correction factors ranged from 0.77 to 1.5 with a median value of 0.97
and a median absolute deviation\footnote{The Median Absolute Deviation
  is defined as MAD=1.4826 $\times$ median\{$|$x$_i-$median(x)$|$\}
  and is a robust estimate of the standard deviation, $\sigma$, of a
  distribution.  The constant of 1.4826 is defined such that
  MAD=$\sigma$ if the random variable $x$ follows a normal
  distribution and the sample is large.} of 0.04.

A subset of the M, L, and T dwarf SL spectra are shown in Figures
1$-$3.  The signal-to-noise ratio (S/N) of the spectra range from
several hundred for the early-type M dwarfs to a few for the faintest
T dwarfs.  Prominent absorption features of H$_2$O, CH$_4$, and NH$_3$
are indicated.  Figure \ref{LLSeq} shows the LL spectra of those
dwarfs in our sample with the highest S/N.  The S/N ratio of the
spectra range from $>$100 for Gl 229A to a few for DENIS J0255$-$4700.
The LL spectra of ultracool dwarfs are relatively featureless at the
resolving power of the IRS.  \citet{2004ApJ...613..567W} found that
the inclusion of the ground state $\Delta \nu$=+1 bands of LiCl at
$\sim$15.8 $\mu$m affects synthetic spectra at the level of a few
percent at $T_{\mathrm{eff}}$ = 1500 K (the approximate
$T_{\mathrm{eff}}$ of DENIS J0255$-$4700).  Given the low S/N of the
LL spectra, and the predicted weakness of the LiCl bands, we cannot
assess whether this band is present in the spectra of ultracool
dwarfs.  We do not discuss the LL spectra further.

\section{Discussion}

To aid the reader in the interpretation of these spectra, a sequence
of model spectra with $T_{\mathrm{eff}}$s ranging from 3800 K down to
600 K in steps of 400 K with a $\log g$=5.0 is shown in Figure
\ref{ModSeq}.  The models with $T_{\mathrm{eff}}$ $\geq$ 2600 K are
AMES-COND models \citep{2001ApJ...556..357A}, the models with 1400 K
$\leq$ $T_{\mathrm{eff}}$ $<$ 2600 K are cloudy models \citep[][M.~S.
Marley et al., in preparation]{2002ApJ...568..335M} and the models
with $T_{\mathrm{eff}}$ $<$ 1400 K are cloudless models.  The spectra
have been smoothed to $R$=90 and resampled onto the wavelength grid of
the IRS spectra.  Also shown are the approximate spectral types
corresponding to each $T_{\mathrm{eff}}$
\citep{2000ApJ...535..965L,2004AJ....127.3516G}.

\begin{figure}
\includegraphics[width=5.5in]{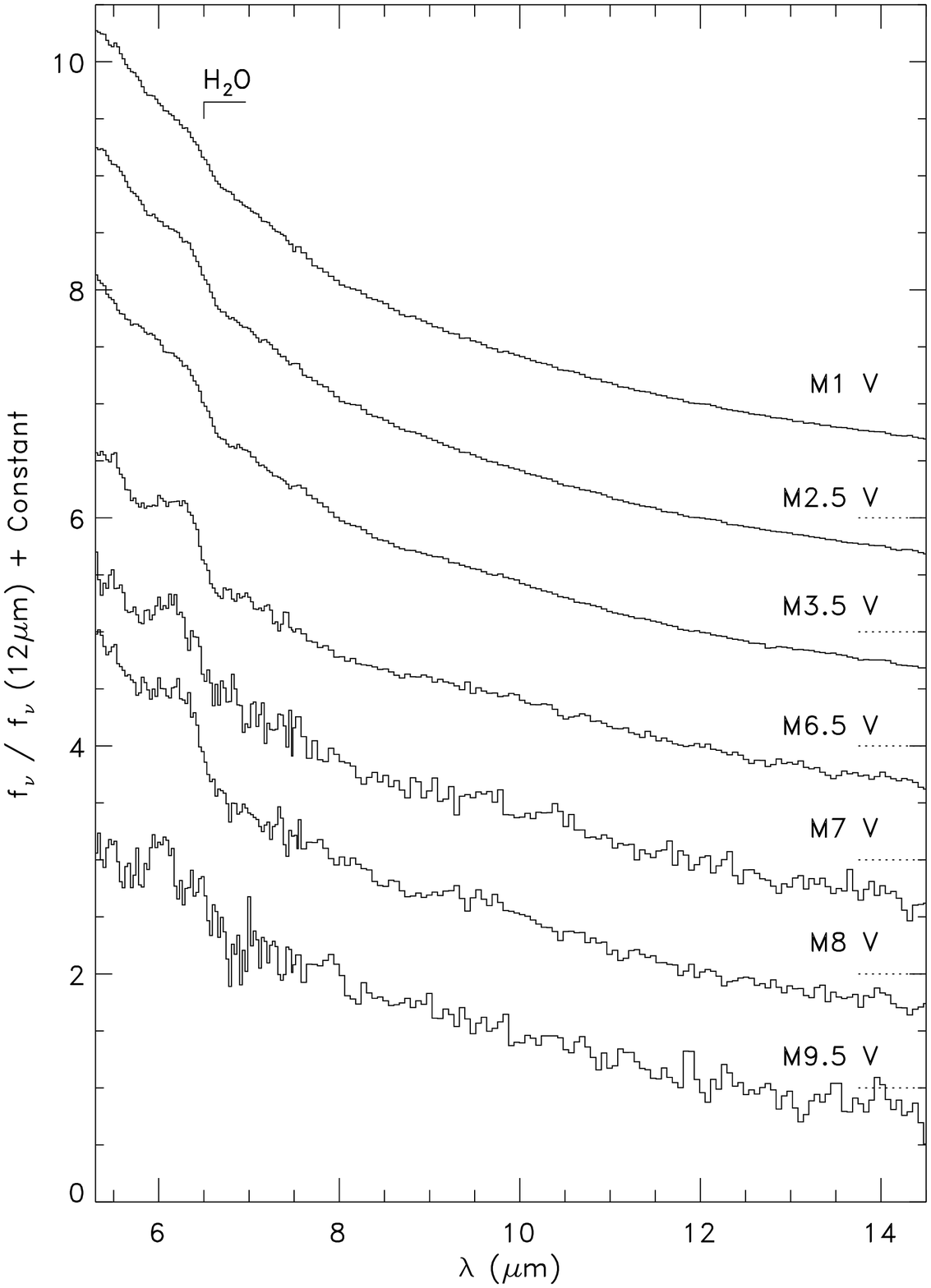}
\caption{\label{Ms}The 5.5$-$14.5 $\mu$m spectra of Gl 229A (M1 V), Gl
  752A (M2.5 V), GJ 1001A (M3.5 V), GJ 1111 (M6.5 V), LHS 3003 (M7 V),
  vB 10 (M8 V), and BRI 0021$-$0214 (M9.5 V).  The spectra have been
  normalized at 12 $\mu$m and offset by constants (\textit{dotted
    lines}); the flux densities of the spectra at 12 $\mu$m are
  872, 519, 36.4, 70.0, 15.4, 17.6, and 4.90 mJy, respectively.}
\end{figure}

\begin{figure}
\includegraphics[width=5.5in]{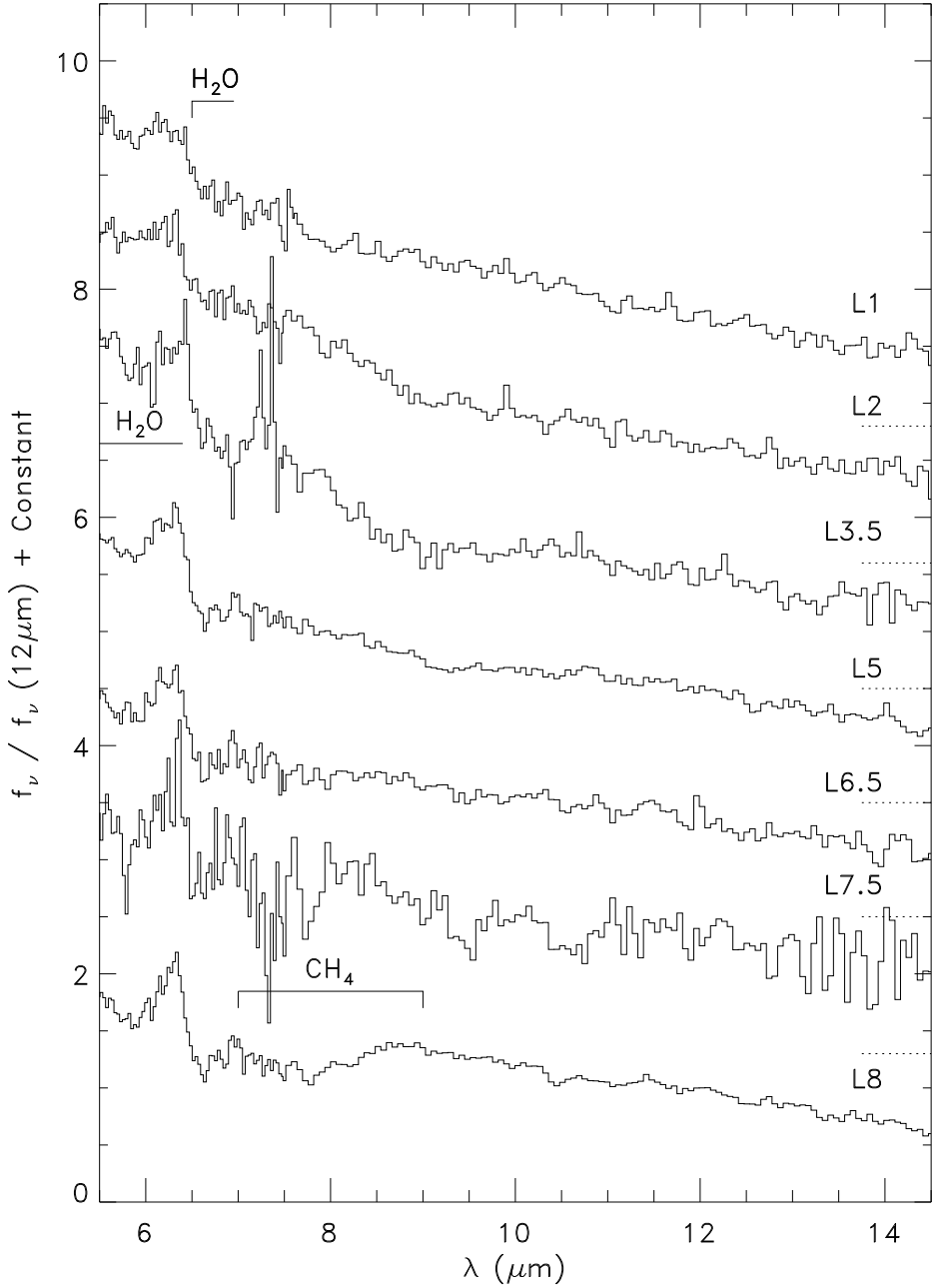}
\caption{\label{Ls}The 5.5$-$14.5 $\mu$m spectra of 2MASS J1439$+$1929
  (L1), Kelu-1AB (L2), 2MASS J0036$+$1821 (L3.5), 2MASS J1507$-$1627
  (L5), 2MASS J1515$+$4847 (L6.5), 2MASS J0825$+$2115 (L7.5), and
  DENIS J0255$-$4700 (L8).  The spectra have been normalized at 12
  $\mu$m and offset by constants (\textit{dotted lines}); the flux
  densities of the spectra at 12 $\mu$m are 2.13, 2.00, 3.57, 4.47,
  2.24, 2.00, and 7.14 mJy, respectively.}
\end{figure}

The spectra of the M and L dwarfs at $\lambda >$ 5 $\mu$m are
dominated almost entirely by absorption features arising from the
$\nu_2$ fundamental band of H$_2$O centered at $\sim$6.27 $\mu$m, and
the 2$\nu_2 - \nu_2$ overtone band centered at $\sim$6.42 $\mu$m.
However due to a combination of the weakness of these features and the
low spectral resolving power of the IRS spectra ($R \approx$90), the
only H$_2$O feature readily identifiable is a ``break'' at $\sim$6.5
$\mu$m.  As the $T_{\mathrm{eff}}$ decreases, this break generally
increases in strength until eventually additional H$_2$O absorption at
longer and shorter wavelengths transforms it into an emission-like
feature in the spectra of the T dwarfs.  In actuality, this emission
feature is a result of a minimum in the H$_2$O opacity which allows
the observer to see deeper, and thus hotter, atmospheric layers.
Counterintuitively, the $T_{\mathrm{eff}}$=2600 K COND model ($\sim$M7
V) shows stronger H$_2$O absorption from 8$-$10 $\mu$m (P. Hauschildt
2005, private communication) than the $T_{\mathrm{eff}}$=2200 K cloudy
model ($\sim$L1).  As can be seen in Figures \ref{Ms} and \ref{Ls}, we
see no evidence of this absorption in the spectra of the late-type M
and early-type L dwarfs.  The $\nu_4$ fundamental band of CH$_4$
centered at $\sim$7.65 $\mu$m first appears in the spectra of the
latest L dwarfs ($T_{\mathrm{eff}}$ $\approx$ 1500 K) and grows in
strength through the T sequence.  The combination of H$_2$O and CH$_4$
absorption from roughly 4 to 9 $\mu$m heavily suppresses the flux at
these wavelengths in the spectra of the T dwarfs.  Finally the $\nu_2$
fundamental band of NH$_3$ centered at $\sim$10.5 $\mu$m appears in
the spectra of the early- to mid-type T dwarfs.  The only clearly
discernible NH$_3$ feature is the double Q-branch feature centered at
10.5 $\mu$m; the double Q-branch is a result of inversion doubling
\citep{1945msms.book.....H}.  Overall, the theoretical spectra provide
a reasonably good match to the mid-infrared spectra of M, L, and T
dwarfs.  A more detailed comparison between the models and the
observations is currently in progress (Cushing et al., in preparation).

We have defined three spectral indices that measure the depths of the
H$_2$O bands at $\sim$6.3 $\mu$m, the 7.65 $\mu$m CH$_4$ band, and the 10.5
$\mu$m NH$_3$ band in the IRS spectra of the M, L, and T dwarfs.
Figure \ref{IndicesExplanation} shows an illustration of the three
spectral indices along with the spectrum of 2MASS J0559$-$1404 (T4.5).
As described above, the only H$_2$O feature readily identifiable in
the IRS spectra of ultracool dwarfs is located at $\sim$6.5 $\mu$m.
We have therefore defined an index that measures the amplitude of the
6.25 $\mu$m peak relative to the two minima on either side.  This
index is given by,

\begin{equation}
\mbox{IRS-H$_2$O} = \frac{f_{6.25}}{0.562 f_{5.80} + 0.474 f_{6.75}},
\end{equation}
where $f_{\lambda_0}$ is the mean flux density in a 0.15 $\mu$m window
centered around $\lambda_0$.  Both the CH$_4$ and NH$_3$ indices are
simple ratios of the flux density in and out of an absorption feature
and are defined as,

\begin{equation}
\mbox{IRS-CH$_4$} = \frac{f_{10.0}}{f_{8.5}},
\end{equation}
\noindent
and 
\begin{equation}
\mbox{IRS-NH$_3$} = \frac{f_{10.0}}{f_{10.8}}
\end{equation}
where $f_{\lambda_0}$ is the mean flux density in a 0.3 $\mu$m window
centered around $\lambda_0$.  The values of the indices computed for
the dwarfs in our sample are shown as a function of spectral type in
Figure \ref{Indices}.  The errors were computed from the uncertainties
in the mean flux densities $f_{\lambda_0}$.  Larger values of a given
index imply stronger absorption.

The IRS-H$_2$O values indicate that, overall, the H$_2$O absorption band
strength increases with increasing spectral type until it saturates in
the T spectral class.  Nevertheless, there also appears to be a plateau
from about $\sim$M7 V to $\sim$L5 indicating that late-type M dwarfs and
early- to mid-type L dwarfs have similar H$_2$O band strengths.  BRI
0021$-$0214 (M9.5 V) and 2MASS J1439$+$1929 (L1) appear to have
anomalously low H$_2$O band strengths.  The variations of the IRS-CH$_4$
and IRS-NH$_3$ values with spectral type are, in contrast, much simpler.
The onset of CH$_4$ absorption occurs at roughly the L/T transition.  A
more precise spectral type cannot be assigned given the coarse
wavelength sampling, low resolving power, and moderate S/N of the IRS
spectra.  Interestingly, the $\nu_3$ fundamental band of CH$_4$ at 3.3
$\mu$m has been detected in the spectra of mid-type L dwarfs
\citep{1998ApJ...502..932O,2001PhDT.......116B,2000ApJ...541L..75N,2005ApJ...623.1115C}.
The absorption cross section of the $\nu_4$ band of CH$_4$ centered at
7.65 $\mu$m is roughly an order of magnitude smaller than that of the
$\nu_3$ band ($T$=1000 K, $P$=1 bar, R. Freedman 2005, private
communication) so it is not surprising that the $\nu_4$ band appears
later in the spectral sequence than the $\nu_3$ band.  Finally, The
IRS-NH$_3$ values also indicate that the onset of NH$_3$ absorption also
occurs near the L/T transition.  Interestingly, the values of both the
IRS-CH$_4$ and IRS-NH$_3$ indices decrease through the L spectral class,
a behavior we discuss in \S3.3.1.

\subsection{Objects of Interest}

As described in the previous section, the mid-infrared spectral
features of M, L, and T dwarfs generally show a smooth variation with
spectral type and are qualitatively well matched by model spectra.
However there are a number of interesting objects that stand out
against this sequence that we discuss in the following section.

\subsubsection{Mid-Type L Dwarfs}

It has been apparent for some time that the atmospheres of L dwarfs are
cloudy.  The formation of these condensate (i.e., dust) clouds in the
atmospheres of ultracool dwarfs has a dramatic impact on their
atmospheric structure ($T/P$ profile) and thus their emergent spectra.
Models that neglect dust formation produce near-infrared colors that are
much bluer than the observations
\citep{2001ApJ...556..357A,2002ApJ...568..335M,2004AJ....127.3553K,2006ApJ...640.1063B}.
However the limited wavelength span of near-infrared spectra has
precluded definitive determinations of either particle size or
condensate composition.  IRS spectra both substantially increase the
wavelength range of L dwarf spectra--allowing for Mie scattering effects
to be constrained--and cover the location of the the 10 $\mu$m silicate
feature.

While the IRS spectra of the early-type L dwarfs and the T dwarfs are
in generally good agreement with the model predictions \citep[M.~C.
Cushing et al., in preparation]{2004ApJS..154..418R}, the spectra of
mid- to late-type L dwarfs differ substantially from the models.
Figure \ref{oddLs} shows a sequence of L dwarfs with spectral types
ranging from L1 to L6.5.  As can been seen, the spectrum of 2MASS
J2224$-$0158 (L4.5) exhibits a prominent plateau from roughly 9 to 11
$\mu$m.  A similar, although weaker, plateau can also be seen in the
spectra of 2MASS J0036$+$1821 (L3.5) and 2MASS J1507$-$1627 (L5).
This feature is also clearly absent in the spectra of L dwarfs with
both earlier and later spectral types.  This plateau is the cause of
the decreasing IRS-CH$_4$ and IRS-NH$_3$ values in the L dwarfs (see
\S3) since these indices are also a measure of the overall spectral
slope in the M and L dwarf spectra.  The broad deviation of the model
from the observed spectra implies that the model is missing or
incorrectly characterizing a continuum opacity source.  Given the good
agreement between model and data at early and late spectral types, a
missing gaseous opacity source with a smooth continuum seems unlikely.
We thus conclude that the most likely explanation for the deviation is
the description of the cloud opacity.

The IRS spectral region of course includes the 10 $\mu$m silicate
feature which arises from the Si-O stretching vibration in silicate
grains.  The spectral shape and importance of the silicate feature
depends on the particle size and composition of the silicate grains.  In
brown dwarf atmospheres the first expected silicate condensate is
forsterite Mg$_2$SiO$_4$ \citep{2002ApJ...577..974L}, at $T \approx
1700$ K ($P$=1 bar) \footnote{Iron-bearing species (e.g.
  olivine, (Mg,Fe)$_2$SiO$_4$) are not expected since iron condenses well
  before the first silicate}. Since Mg and Si have approximately equal
abundances in a solar composition atmosphere, the condensation of
forsterite leaves substantial silicon, present as SiO, in the gas phase.
In equilibrium, at temperatures about 50 to 100 K cooler than the
forsterite condensation temperature, the gaseous SiO reacts with the
forsterite to form enstatite, MgSiO$_3$ \citep{2002ApJ...577..974L}.
Since the precise vertical distribution of silicate species depends upon
the interplay of the atmospheric dynamics and chemistry, the models
(M.~S.  Marley et al., in preparation) do not attempt to capture those
details.  Instead all of the silicate condensates are assumed to be
forsterite--since it condenses first--and the optical properties of
forsterite are employed in the calculation of the Mie absorption and
scattering efficiencies.

In addition to composition, the cloud spectral properties are sensitive
to particle size.  There is likely a range of particle sizes ranging
from very small, recently condensed grains, to larger grains that have
grown by accumulation \citep{2001ApJ...556..872A,2004A&A...414..335W}.
The atmosphere model includes a calculation of turbulent diffusion and
particle sedimentation to compute a mean particle size
\citep{2001ApJ...556..872A} assuming a log-normal size distribution with
fixed width $\sigma = 2$.  The Ackerman \& Marley cloud model predicts
submicron particle sizes high above the condensation layer (as do
\citet{2004A&A...414..335W}), but these small particles do not provide
enough opacity to produce a detectable effect on the model spectra.  In
the optically-thick cloud the model predicts mean particles sizes of 5
to 10 $\mu$m and larger.  Such a population of particles is too
large to produce a ten micron silicate feature.

Figure \ref{Grains} compares the absorption efficiency of silicate
grains of various sizes, composition, and crystal structures to the
spectrum of 2MASS J2224$-$0158 (L4.5).  For each species the quantity
$Q_{\mathrm{abs}}/a$, or Mie absorption efficiency divided by particle
radius, is shown.  This is the relevant quantity since, all else being
equal, the total cloud absorption optical depth is proportional to this
quantity \citep{2000fgpc.conf..152M}.  The middle panel of Figure
\ref{Grains} leads us to conclude that the mismatch between the models
and data may arise from a population of silicate grains that is not
captured by the cloud model.  The large grain sizes computed by the
cloud model ($\sim$10 $\mu$m) tend to have relatively flat absorption
spectra (\textit{dashed lines}) across the IRS spectral range.  Only
grains smaller than about 2 $\mu$m in radius show the classic 10-$\mu$m
silicate feature \citep{1994ApJ...425..274H} which suggests that the
cloud model does not produce enough small grains.  In addition, the
combined width of the enstatite, whose opacity is currently not included
in the atmosphere models, and forsterite features is a better match to
the width of the plateau in the spectrum of 2MASS J2224$-$0158.

Furthermore the model employs amorphous silicate optical properties.  It
is possible, especially at the higher pressures found in brown dwarf
atmospheres, that the grains are crystalline, not amorphous.  Indeed
laboratory solar-composition condensation experiments at relevant
pressures produce crystalline, not amorphous, silicates
\citep{2004LPI....35.1726T}.  Crystalline grains (lower panel of Figure
\ref{Grains}) can have larger--and spectrally richer--absorption cross
sections.  The strongest absorption feature of crystalline enstatite in
the IRS wavelength range occurs at $\sim$9.17 $\mu$m.  The weak
absorption feature in the IRS spectrum of 2MASS J1507$-$1627 (L5; see
Figure \ref{oddLs}) at the same wavelength may therefore be carried by
crystalline enstatite.  However, higher S/N ratio spectra would be
required to confirm this tentative identification.

Finally we note that \citet{helling06} predict that non-equilibrium
effects will lead to the condensation of quartz ($\rm SiO_2$) grains
within the silicate cloud.  Quartz absorption begins somewhat bluer than
that of enstatite and, given small enough particles sizes, also might
add to the spectral flattening seen in Figures 8 and 9.  Along with IRS
observations of more mid L dwarfs, detailed cloud modeling considering a
range of cloud sizes and compositions will be required to fully
constrain the particular species, particle sizes, and crystallinity
present in the silicate cloud.

\subsubsection{Gl 337CD \& SDSS J0423$-$0414AB}

Gl 337CD was discovered by \citet{2001AJ....122.1989W} and later
resolved into a near equal-magnitude ($K_S$ flux ratio of
0.93$\pm$0.10) binary separated by 0$\farcs$53 by
\citet{2005AJ....129.2849B}.  Its unresolved near-infrared spectrum
exhibits weak CH$_4$ absorption \citep{2003ApJ...596..561M} resulting
in a near-infrared spectral type of T0 \citep{2006ApJ...637.1067B}
while its unresolved red-optical spectrum has been typed as L8
\citep{2001AJ....122.1989W}.  The absolute $K_s$ magnitudes of the
components provide little constraint on the individual spectral types
of the two objects since the $M_{Ks}$ values are consistent with a
broad range of types from late-type L dwarfs through mid-type T
dwarfs. Given the composite optical and near-infrared spectral types
of L8 and T0, respectively, the pair is likely comprised of a late-L
and early- to mid-T dwarf.  SDSS J0423$-$0414AB (hereafter SDSS
0423AB) was discovered by \citet{2002ApJ...564..466G} and subsequently
resolved into a binary separated by 0$\farcs$16 by
\citet{2005ApJ...634L.177B}.  It was also classified as T0 based on
the presence of weak CH$_4$ absorption in its near-infrared spectrum
\citep{2002ApJ...564..466G,2006ApJ...637.1067B} and has an unresolved
optical spectral type of L7.5 \citep[][J.~D. Kirkpatrick, in
preparation]{2003AJ....126.2421C}.  \citet{2005ApJ...634L.177B} found
that a hybrid spectrum composed of an L6.5 and T2 dwarf provides an
excellent match to the unresolved near-infrared spectrum of SDSS
0423AB.

Figure \ref{Gl337CD} shows the IRS spectra of SDSS 0423AB and Gl
337CD. Although the spectra have almost identical (unresolved) optical
and near-infrared spectral types, their mid-infrared spectra look
markedly different.  In particular, the CH$_4$ band centered at 7.65
$\mu$m is much stronger in the spectrum of Gl 337CD (see also figure
\ref{Indices}).  Also shown are composite L5$+$T2 and L8$+$T4.5
spectra (red lines) constructed after scaling the spectra of 2MASS
J1507$-$1627 (L5), SDSS J1254$-$0122 (T2), DENIS J0255$-$4700 (L8),
and 2MASS J0559$-$1404 (T4.5) to appear as if they were at a common
distance.  Although \citet{2005ApJ...634L.177B} found that an
L6.5$+$T2 composite near-infrared spectrum was the best match to that
of SDSS 0423AB, we used an L5 dwarf since our sample lacks an L6 dwarf
with a measured trigonometric parallax.  Nevertheless, the agreement
between the data and composite spectra is quite good.  The weak CH$_4$
band in SDSS 0423AB is therefore a result of the intrinsically
brighter L6.5 dwarf, which lacks CH$_4$ absorption, veiling the CH$_4$
band in the spectrum of the T2 dwarf.  In contrast, both components of
Gl337CD (L8, T4.5) exhibit CH$_4$ absorption resulting in a prominent
CH$_4$ band in its unresolved spectrum.

\subsubsection{2MASS J0937$+$2931}

2MASS J0937$+$2931 (hereafter 2MASS 0937) is the archetypal peculiar T
dwarf.  It is classified as a T6p \citep{2006ApJ...637.1067B} because
it exhibits a number of spectral peculiarities including an enhanced
emission peak at 1.05 $\mu$m, weak $J$-band \ion{K}{1} lines, and a
heavily suppressed $K$-band spectrum
\citep{2002ApJ...564..421B,2004AJ....127.3553K}.  All of these
spectral features are indicative of high pressure (high surface
gravity) and/or low-metallicity atmospheres.  In particular, the
suppression of the $K$-band is a result of collision-induced H$_2$
1$-$0 dipole absorption (CIA H$_2$) centered at 2.4 $\mu$m
\citep{2002A&A...390..779B,2004AJ....127.3553K} which is enhanced in
such environments.  Indeed, \citet{burgasser05b} have shown that a
synthetic spectrum with a moderately low metallicity ($-$0.1 $\leq$
[M/H] $\leq$ $-$0.4) and high surface gravity (5.0 $\leq$ $\log g$
$\leq$ 5.5) is required to adequately fit its 0.7$-$2.5 $\mu$m
spectrum.

Figure \ref{T6Comp} shows the IRS spectrum of 2MASS 0937 along with the
spectrum of SDSS J1624$+$0029 (T6; hereafter SDSS 1624).  The spectrum of
SDSS 1624 has been scaled by the ratio of the distances of the two
objects to adjust its flux to the level that which would be observed if
it were at the distance of 2MASS 0937.  The spectrum of 2MASS 0937
appears significantly depressed shortward of $\sim$7.5 $\mu$m relative
to the spectrum of SDSS 1624.  Although we tentatively ascribe this
behavior to the subsolar metallicity/high surface gravity of 2MASS 0937,
we caution that additional high S/N IRS observations of late-type T
dwarfs will be required to confirm that the mid-infrared spectrum of
2MASS 0937 is truly distinct from typical T dwarfs.

\section{Spectral Energy Distributions \& Bolometric Fluxes}

\subsection{Spectral Energy Distributions}

Figure \ref{SEDSeq} shows the 0.6$-$14.5 $\mu$m spectra of GJ 1111 (M6.5
V), 2MASS J1507$-$1627 (L5) and 2MASS J0559$-$1404 (T4.5).  The
red-optical spectra are from \citet{1991ApJS...77..417K},
\citep{2000AJ....119..928F}, and \citet{2003ApJ...594..510B} and the
near-infrared spectra are from \citet{2005ApJ...623.1115C} and J.~T.
Rayner et al. (in preparation).  The changes in the spectral morphology
across the MLT sequence illustrate all of the major chemical transitions
expected to occur in the atmospheres of ultracool dwarfs
\citep{1996ApJ...472L..37F,1999ApJ...519..793L,1999ApJ...512..843B,2002Icar..155..393L}.

In the atmospheres of M dwarfs (2400 K $\lesssim$ $T_{\mathrm{eff}}$
$\lesssim$ 3800 K), C, N, and O are found primarily in CO, N$_2$, and
H$_2$O.  The spectral morphology of M dwarfs is therefore shaped
primarily by H$_2$O absorption bands although TiO and VO bands dominate
in the optical. The $\Delta \nu = +2$ CO bands at $\lambda \gtrsim$ 2.29
$\mu$m and absorption lines of refractory species such as Al, Mg, Fe,
and Ca are also weakly present.  Since N$_2$ is a homonuclear molecule,
it cannot radiate in the dipole approximation and therefore shows no
detectable spectral signatures in the spectra of M and L dwarfs
(although N$_2$ can in principle absorb via collisions with H$_2$
molecules akin to the CIA H$_2$ opacity \citep{1986ApJ...303..495B}).

As $T_{\mathrm{eff}}$ approaches 2400 K, condensates begin forming in
the atmospheres of ultracool dwarfs.  In particular, titanium- and
vanadium-bearing condensates form resulting in a loss of TiO and VO
from the gas \citep{2002ApJ...577..974L}; the weakening and eventual
loss of the TiO and VO bands mark the transition to the L spectral
class (1400 $\lesssim$ $T_{\mathrm{eff}}$ $\lesssim$ 2400 K).  With
the loss of the TiO and VO bands, the resonant \ion{K}{1} doublet
becomes very prominent in the spectra of L dwarfs and eventually comes
to define the continuum hundreds of Angstroms from line center.  In
the near-infrared, the H$_2$O and CO bands strengthen with decreasing
$T_{\mathrm{eff}}$.  Additional condensates, most notably Ca-, Al-,
Fe-, Mg- and Si-bearing species, also form and affect the emergent
spectra of L dwarfs by altering the temperature/pressure profile of
the atmosphere and contributing their own opacities.  The near-infrared
colors of the L dwarfs become progressively redder due to the
formation of these condensates.  At the lower end of this
$T_{\mathrm{eff}}$ range, CH$_4$ becomes the dominate carbon-bearing
species in the upper, coolest layers of the atmosphere, since
CO/CH$_4 <$ 1 for $T\lesssim$ 1100 K at $P$=1 bar
\citep{2002Icar..155..393L}.  Indeed the $\nu_3$ fundamental band of
CH$_4$ at 3.3 $\mu$m, which is $\sim$100 times stronger than the
combination and overtone bands in the near-infrared and $\sim$10 times
stronger than the $\nu_4$ fundamental band in the mid-infrared, can be
seen in the spectra of mid- to late-type L dwarfs.

As $T_{\mathrm{eff}}$ continues to decrease, CH$_4$ becomes ever more
dominant over CO; the appearance of the CH$_4$ overtone and
combination bands in the near-infrared signal the transition to the T
spectral class (600 K $\lesssim$ $T_{\mathrm{eff}}$ $\lesssim$ 1400
K).  The condensates which help shape the spectral morphology of the L
dwarfs form well below the observable photosphere in T dwarfs
resulting in a relatively condensate-free atmosphere.  The strong
H$_2$O and CH$_4$ bands carve the near-infrared spectra of T dwarfs up
into narrow bands centered at 1.25, 1.6, and 2.2 $\mu$m.  Finally
NH$_3$ becomes the dominant nitrogen-bearing gas since N$_2$/NH$_3 <$ 1
for $T\lesssim700$ K at $P$=1 bar \citep{2002Icar..155..393L} and
consequently the $\nu_2$ fundamental band of NH$_3$ at $\sim$10.5
$\mu$m is present in the spectra of T dwarfs.

\subsection{Bolometric Luminosities}

The effective temperatures of ultracool dwarfs are typically determined
by combining observed bolometric luminosities with theoretical radii
\citep{2001ApJ...548..908L,2002AJ....124.1170D,2004AJ....127.3516G}.
The bolometric luminosities are measured using absolutely
flux-calibrated optical and near-infrared spectra, $L'$-band (and
sometimes $M'$-band) photometry to account for the flux between
$\sim$2.5 and $\sim$4 $\mu$m, and a Rayleigh-Jeans tail at $\lambda
\gtrsim$ 4 $\mu$m.  Although \citet{2005ApJ...623.1115C} have shown that
$L'$-band photometry can be used as a substitute for spectroscopy from
2.9 to 4.1 $\mu$m for spectral types ranging from M1 to T4.5, the
assumption of a Rayleigh-Jeans tail at $\lambda \gtrsim$ 4 $\mu$m has
never been tested observationally.  The IRS spectra are ideal for this
purpose.

Twelve of the dwarfs in our sample have both published absolutely flux
calibrated 0.6$-$4.1 $\mu$m spectra \citep{2005ApJ...623.1115C} and IRS
spectra.  In order to construct spectra suitable for integration over
all wavelengths, we modified each spectrum by linearly interpolating
from zero flux at zero wavelength to its bluest wavelength and removing
the gaps in wavelength coverage from 1.85 to 2.6 $\mu$m and 4.1 to 5.5
$\mu$m by linear interpolation between the flux densities at the gap
edges.  Finally we extend a Rayleigh-Jeans tail from the reddest
wavelength of each spectrum to infinity.  In order to perform as
accurate a comparison as possible with the results of
\citet{2004AJ....127.3516G}, we use the same parallaxes and assume
$M_{\mathrm{bol}\odot}$ = $+$4.75. The results are listed Table 2 along
with the values derived by \citet{2004AJ....127.3516G}.  We find that
the bolometric magnitudes of the twelve dwarfs agree within the errors
except for 2MASS J1439$+$1929 (L1) which is discrepant by just over
1-$\sigma$.  The $L_{\mathrm{bol}}$s, and thus the $T_{\mathrm{eff}}$s,
of the ultracool dwarfs with spectral types ranging from M1 V to T4.5
presented by \citet{2004AJ....127.3516G} are therefore robust against
any systematic errors introduced using photometry and a Rayleigh-Jeans
to account for the flux between from 2.5 and 15 $\mu$m.

\section{Summary}

We have presented a spectroscopic sequence of M, L, and T dwarfs from
5.5 to 38 $\mu$m at $R$ $\approx$ 90 obtained with the IRS onboard the
\textit{Spitzer Space Telescope}.  The spectra exhibit prominent
absorption bands of H$_2$O, CH$_4$, NH$_3$ and are relatively
featureless at $\lambda \gtrsim$ 15 $\mu$m.  H$_2$O absorption features
are present throughout the MLT sequence while the CH$_4$ and NH$_3$
bands first appear at roughly the L/T transition.  We tentatively
ascribe a plateau in the spectra of a number of mid-type L dwarfs from 9
to 11 $\mu$m to the effects of a population of small silicate grains,
likely lying above the main cloud deck, that are not predicted in
current cloud models.  The spectrum of the mildly metal-poor, high
surface gravity, T dwarf 2MASS J0937$+$2931 (T6p) is suppressed from
5.5$-$7.5 $\mu$m relative to typical T6 dwarfs indicating that
mid-infrared spectroscopy may be a useful probe of surface gravity
and/or metallicity variations.  Finally, we computed bolometric
magnitudes for 12 of the dwarfs in our sample with previously published
0.6$-$4.1 $\mu$m spectra and find good agreement with the values of
Golimowski et al. who use $L'$- and $M'$-band photometry and to account
for the flux emitted at $\lambda >$ 2.5 $\mu$m.

\acknowledgments

We thank Brian Patten for providing the IRAC Band 4 observations in
advance of publication, Peter Hauschildt for providing the AMES-COND
synthetic spectra, John Rayner for providing the near-infrared
spectrum of GJ 1111 in advance of publication, and Richard Freedman,
Katherina Lodders, Diane Wooden, Kelle Cruz, J.D. Smith, and William
Vacca for useful discussions.  This publication makes use of data from
the Two Micron All Sky Survey, which is a joint project of the
University of Massachusetts and the Infrared Processing and Analysis
Center, and funded by the National Aeronautics and Space
Administration and the National Science Foundation, the SIMBAD
database, operated at CDS, Strasbourg, France, NASA's Astrophysics
Data System Bibliographic Services, the M, L, and T dwarf compendium
housed at DwarfArchives.org and maintained by Chris Gelino, Davy
Kirkpatrick, and Adam Burgasser, and the NASA/ IPAC Infrared Science
Archive, which is operated by the Jet Propulsion Laboratory,
California Institute of Technology, under contract with the National
Aeronautics and Space Administration.  This work is based (in part) on
observations made with the Spitzer Space Telescope, which is operated
by the Jet Propulsion Laboratory, California Institute of Technology
under a contract with NASA and is supported (in part) by the United
States Department of Energy under contract W-7405-ENG-36, and NASA
through the Spitzer Space Telescope Fellowship Program, through a
contract issued by the Jet Propulsion Laboratory, California Institute
of Technology under a contract with NASA.  T.L.R acknowledges the
support of NASA's Science Mission Directorate.

\bibliographystyle{apj}
\bibliography{ref,tmp}

\clearpage

\clearpage

\clearpage

\begin{figure}
\includegraphics[width=5.5in]{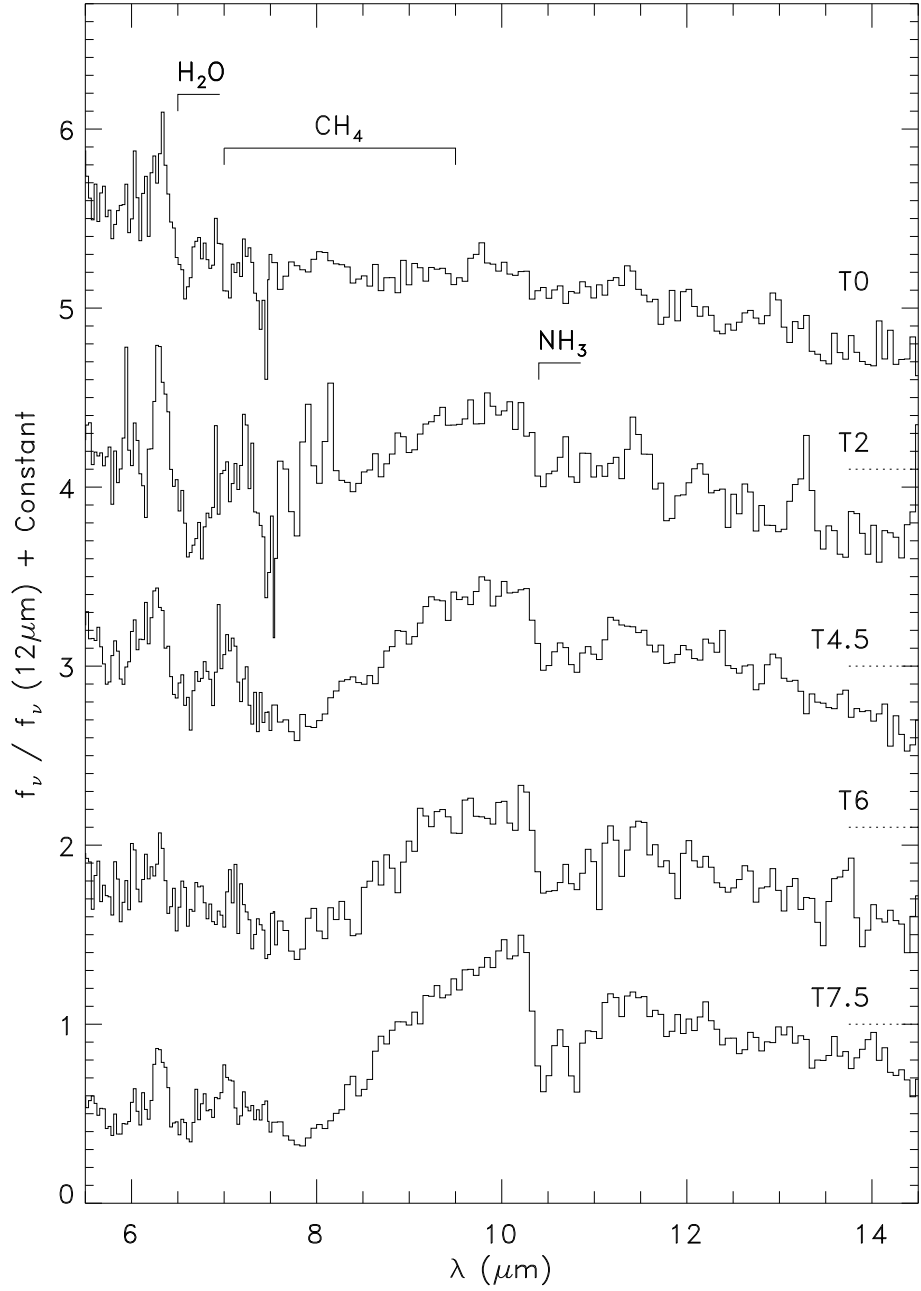}
\caption{\label{Ts}The 5.5$-$14.5 $\mu$m spectra of SDSS
  J0423$-$0414AB (T0), SDSS J1254$-$0122 (T2), 2MASS J0559$-$1404
  (T4.5), SDSS J1624$+$0029 (T6), and Gl 570D (T7.5).  The spectra
  have been normalized at 12 $\mu$m and offset by constants
  (\textit{dotted lines}); the flux densities of the spectra at 12
  $\mu$m are 2.27, 1.40, 2.10, 0.775, and 1.82 mJy, respectively.}

\end{figure}

\clearpage

\begin{figure}
\includegraphics[width=5.5in]{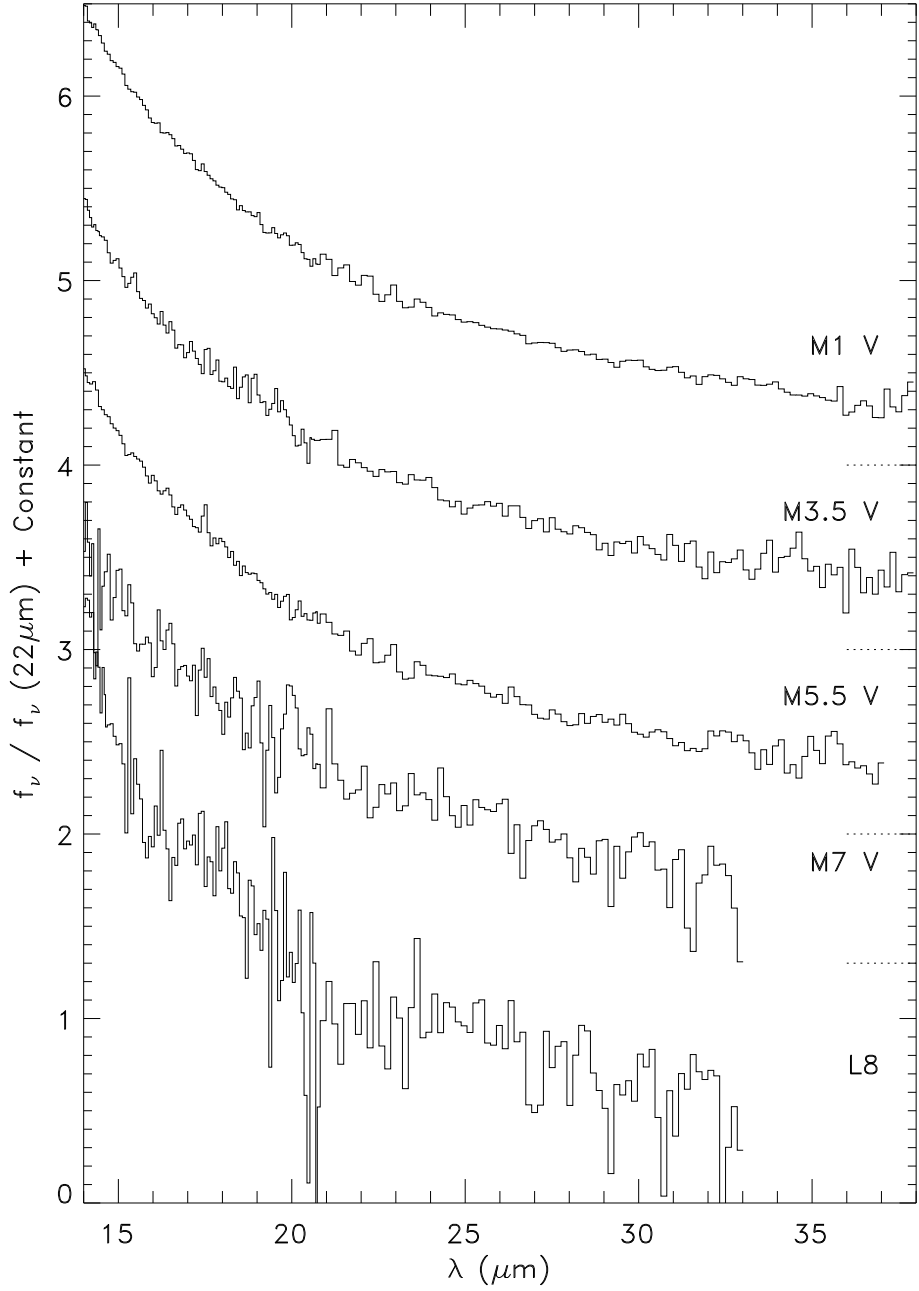}
\caption{\label{LLSeq}The 15.0$-$38.0 $\mu$m spectra of Gl 229A (M1
  V), GJ 1001A (M3.5 V), Gl 65AB (M5.5 V), LHS 3003 (M7 V), and DENIS
  J0255$-$4700 (L8).  The spectra have been normalized at 22 $\mu$m
  and offset by constants (\textit{dotted lines}); the flux densities
  at 22 $\mu$m are 242, 11.5, 91.0, 4.87, and 1.57 mJy, respectively.  The
  longest wavelengths have been removed from some of the spectra due
  to a low signal-to-noise ratio.}
\end{figure}

\clearpage

\begin{figure}
\includegraphics[width=5.5in]{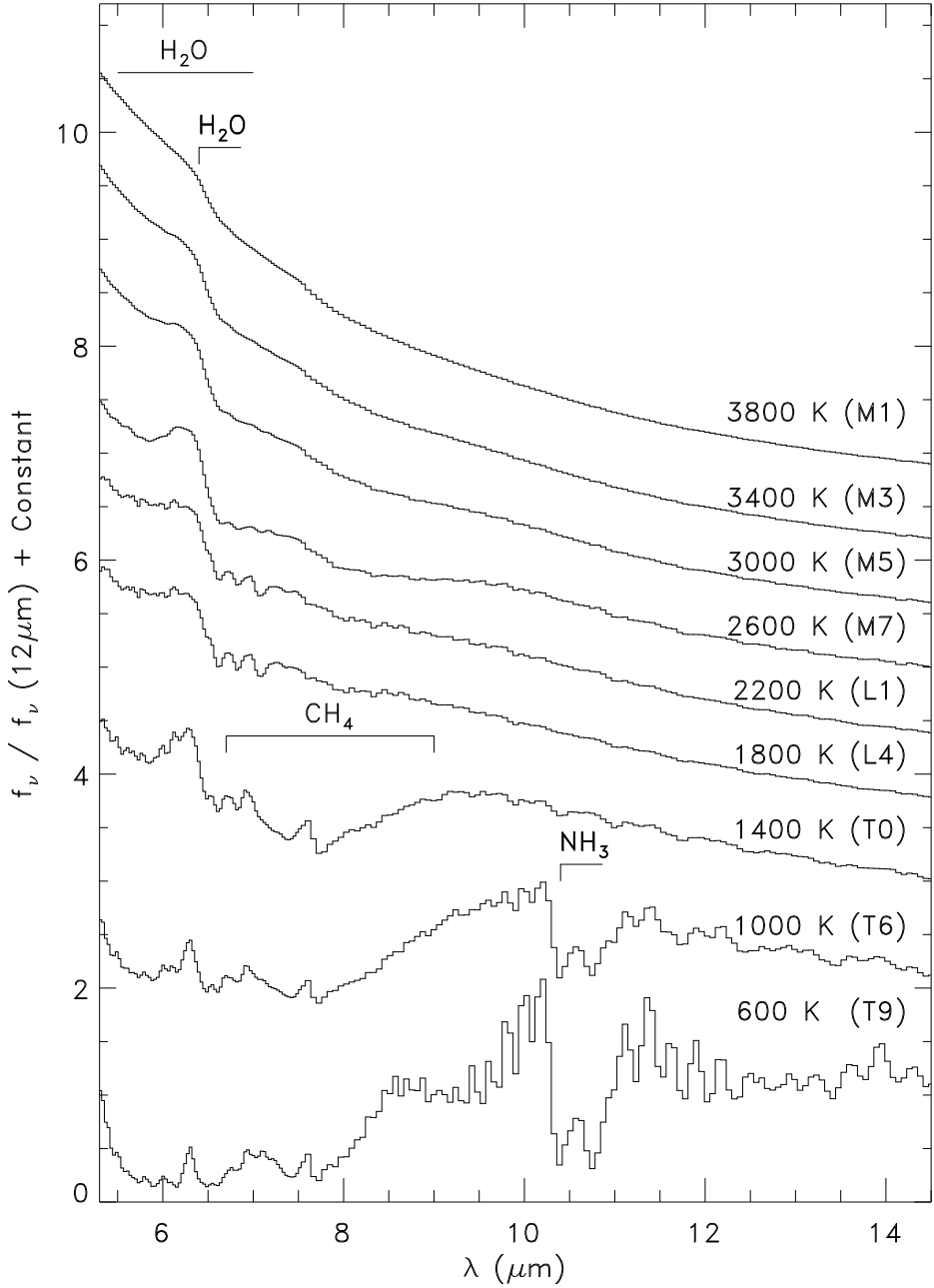}
\caption{\label{ModSeq}Model sequence from $T$=3800 K to $T$=600 K in
  steps of 400 K.  The models with $T_{\mathrm{eff}}$ $\geq$ 2600 K
  are AMES-COND models \citep{2001ApJ...556..357A}, the models with
  1400 K $\leq$ $T_{\mathrm{eff}}$ $<$ 2600 K are cloudy models
  \citep[][M.~S.  Marley et al., in preparation]{2002ApJ...568..335M}
  and the models with $T_{\mathrm{eff}}$ $<$ 1400 K are cloudless
  models.  The spectra have been smoothed to $R$=90 and resampled onto
  the wavelength grid of the IRS spectra.  The approximate spectral
  types corresponding to the effective temperatures are from
  \citet{2000ApJ...535..965L} and \citet{2004AJ....127.3516G}}
\end{figure}

\clearpage

\begin{figure}
\includegraphics[width=5.5in]{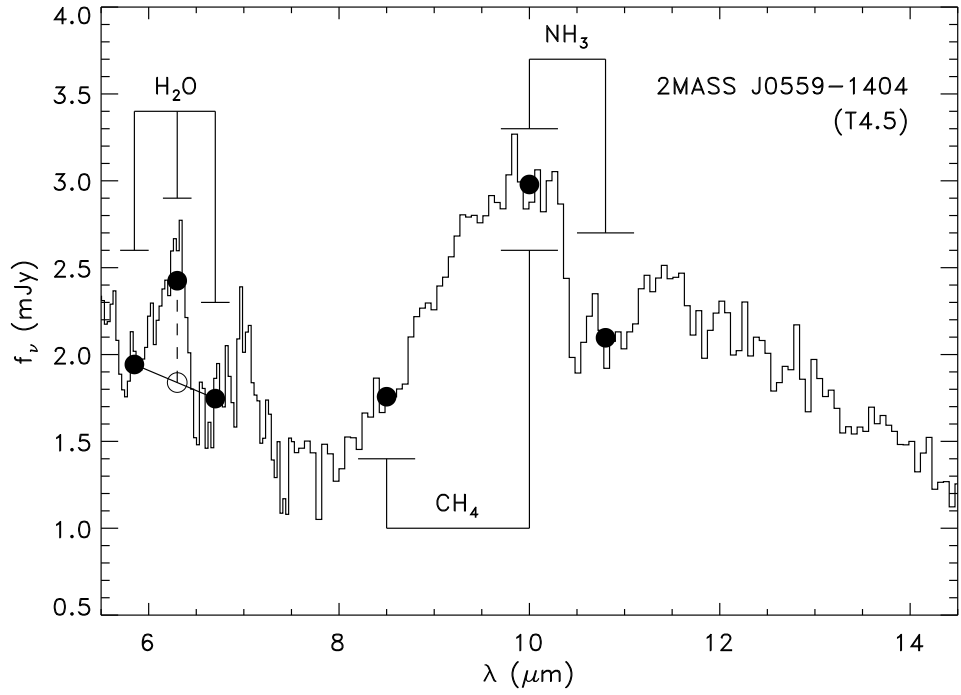}
\caption{\label{IndicesExplanation}Illustration of the H$_2$O, CH$_4$,
  and NH$_3$ spectral indices with the spectrum of 2MASS J0559$-$1404
  (T4.5).}
\end{figure}

\clearpage

\begin{figure}
\includegraphics[width=5.5in]{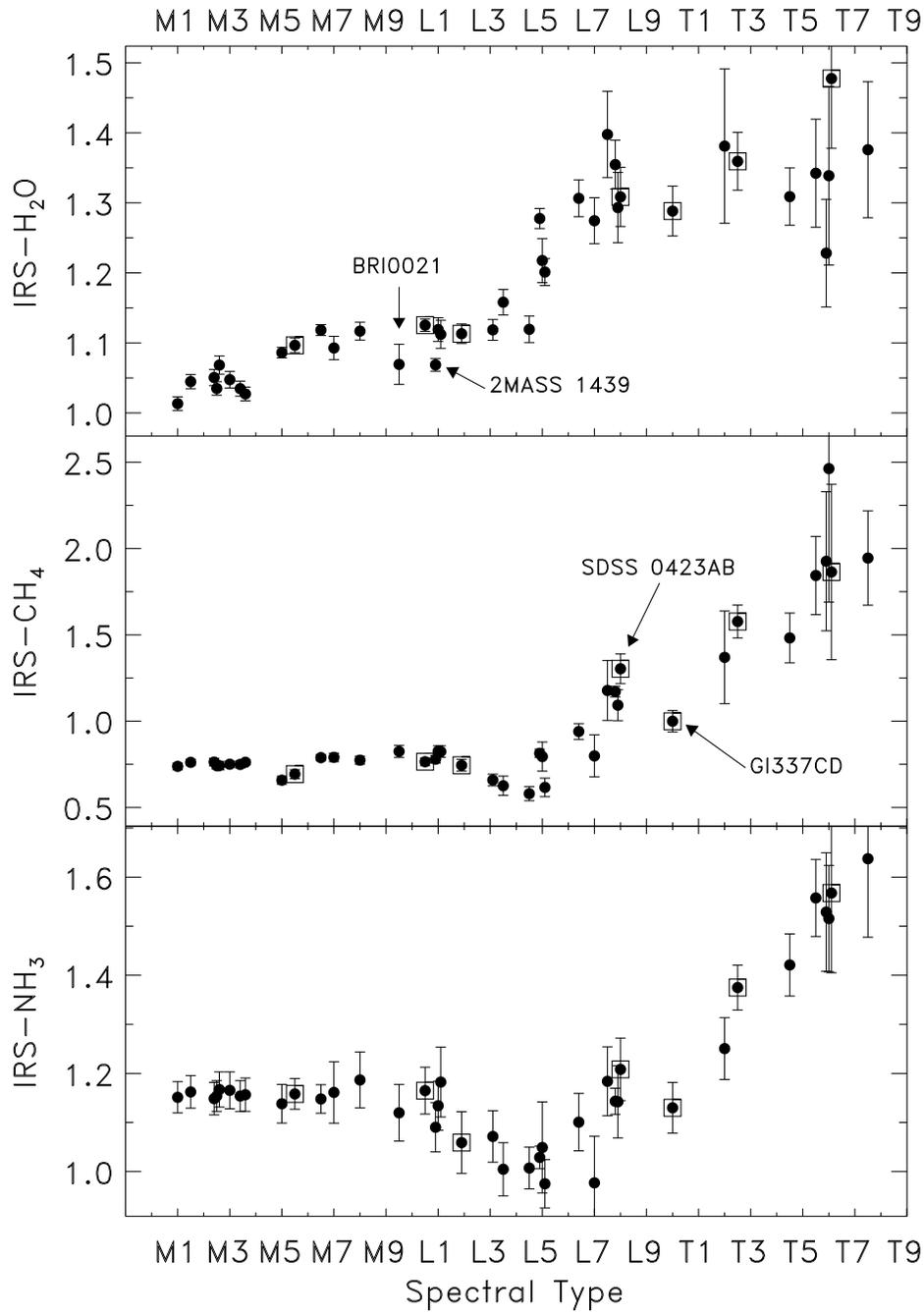}
\caption{\label{Indices}The IRS-H$_2$O, IRS-CH$_4$, and IRS-NH$_3$
  spectral indices of the dwarfs in our sample as a function of
  spectral type.  Individual objects discussed in the text are
  indicated and known binaries are marked with open squares.}
\end{figure}

\clearpage

\begin{figure}
\includegraphics[width=5.5in]{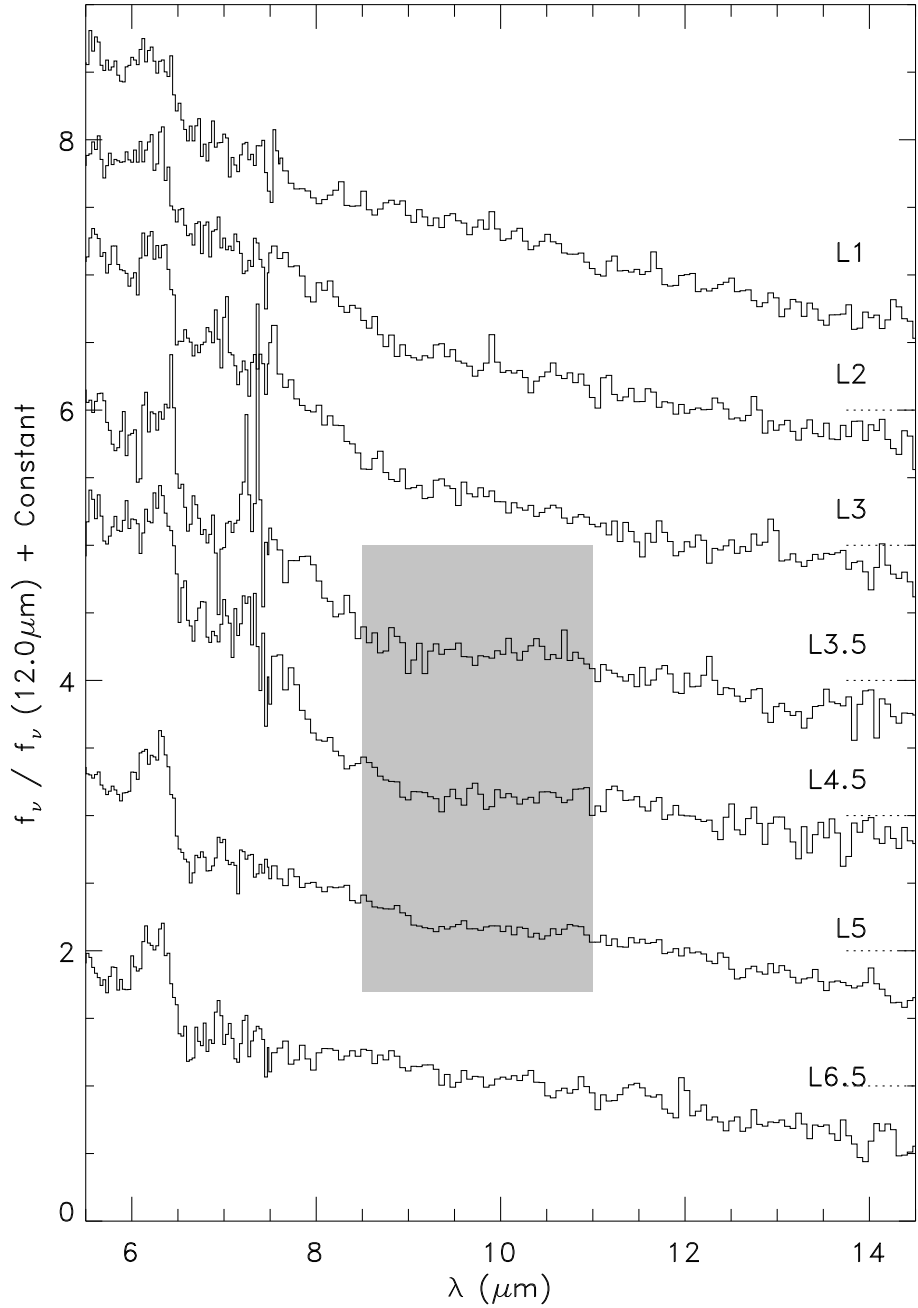}
\caption{\label{oddLs}The IRS spectra of 2MASS J1439$+$1929 (L1),
  Kelu-1AB (L2), 2MASS J1506$+$1321 (L3), 2MASS J0036$+$1821 (L3.5),
  2MASS J2224$-$0158 (L4.5), 2MASS J1507$-$1627 (L5), and 2MASS
  J1515$+$4847 (L6.5).  The spectra have been normalized at 12 $\mu$m
  and offset by constants (\textit{dotted lines}); the flux densities of
  the spectra at 12 $\mu$m are 2.13, 2.00, 1.94, 3.57, 1.64, 4.47, and
  2.24 mJy, respectively.  The grey box indicates the wavelength range
  (9 to 11 $\mu$m) and objects that exhibit the plateau.}
\end{figure}

\clearpage

\begin{figure}
\includegraphics[width=5.5in]{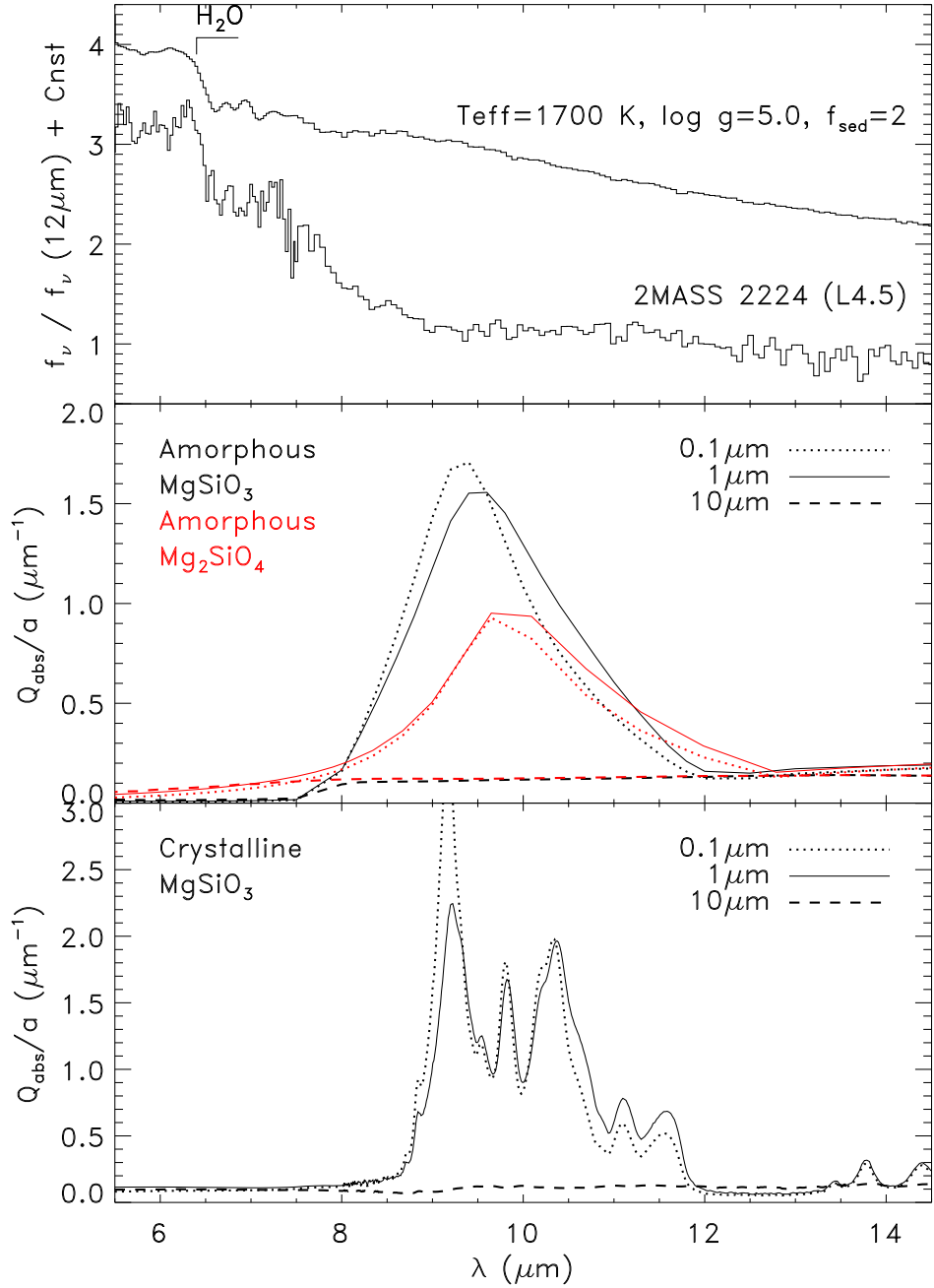}
\caption{\label{Grains}\textit{Top}: Spectrum of 2MASS 2224 (L4.5) and
  a model ($T_{\mathrm{eff}}$=1900 K, $\log g$=5.0,
  $f_{\mathrm{sed}}$=2) appropriate for an L4.5 dwarf from M.S.
  Marley et al. (in preparation).  \textit{Middle}: Optical absorption
  ($Q_{abs}$/a) for amorphous enstatite (MgSiO$_3$; black) and
  forsterite (Mg$_2$SiO$_4$; red) for three different
  particle sizes 0.1 $\mu$m (\textit{dotted lines}), 1 $\mu$m
  (\textit{solid lines}), and 10 $\mu$m (\textit{dashed lines}).
  \textit{Bottom}: Optical absorption ($Q_{abs}$/a) for crystalline
  enstatite (MgSiO$_3$) for three different particle sizes 0.1 $\mu$m
  (\textit{dotted lines}), 1 $\mu$m (\textit{solid lines}), and 10
  $\mu$m (\textit{dashed lines}).  }

%\caption{\label{Grains}\textit{Bottom}: Optical absorption
% ($Q_{abs}$/a) for amorphous enstatite (MgSiO$_3$; black) and
% forsterite (Mg$_2$SiO$_4$; \textit{grey}) for three different
% particle sizes 0.1 $\mu$m (\textit{dotted lines}), 1 $\mu$m
% (\textit{solid lines}), and 10 $\mu$m (\textit{dashed lines}).
% \textit{Middle}: Optical absorption ($Q_{abs}$/a) for crystalline
% enstatite (MgSiO$_3$) for three different particle sizes 0.1 $\mu$m
% (\textit{dotted lines}), 1 $\mu$m (\textit{solid lines}), and 10
% $\mu$m (\textit{dashed lines}).  \textit{Top}: Spectrum of 2MASS
% 2224 (L4.5) and the best fitting model ($T_{\mathrm{eff}}$=1900 K,
% $\log g$=5.0, $f_{\mathrm{sed}}$=2) from M.S. Marley et al. (in
% preparation).  \textit{offset, and check Teff, reverse bottom to
%   top, labels for enstatite and forsterite}}
\end{figure}

\clearpage

\begin{figure}
\includegraphics[width=5.5in,angle=0]{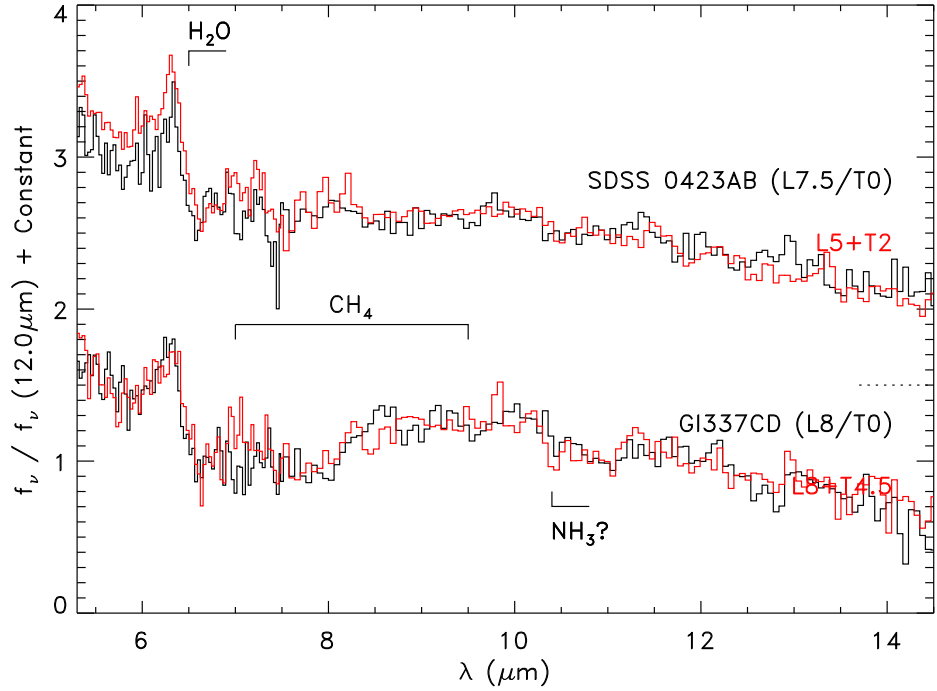}
\caption{\label{Gl337CD}The IRS spectra of SDSS 0423$-$0414AB and Gl
  337CD (\textit{black lines}). Their unresolved optical/near-infrared
  spectral types are L7.5/T0 and L8/T0, respectively.  Also shown are
  composite L5$+$T2 and L8$+$T4.5 spectra (\textit{red})
  constructed from the spectra of 2MASS J1507$-$1627 (L5), SDSS
  J1254$-$0122 (T2), Gl 584C (L8), and 2MASS J0559$-$1404 (T4.5).  The
  spectra have been normalized to unity at 12 $\mu$m and offset by
  constants (\textit{dotted lines}).}
\end{figure}

\clearpage

\begin{figure}
\includegraphics[width=5.5in]{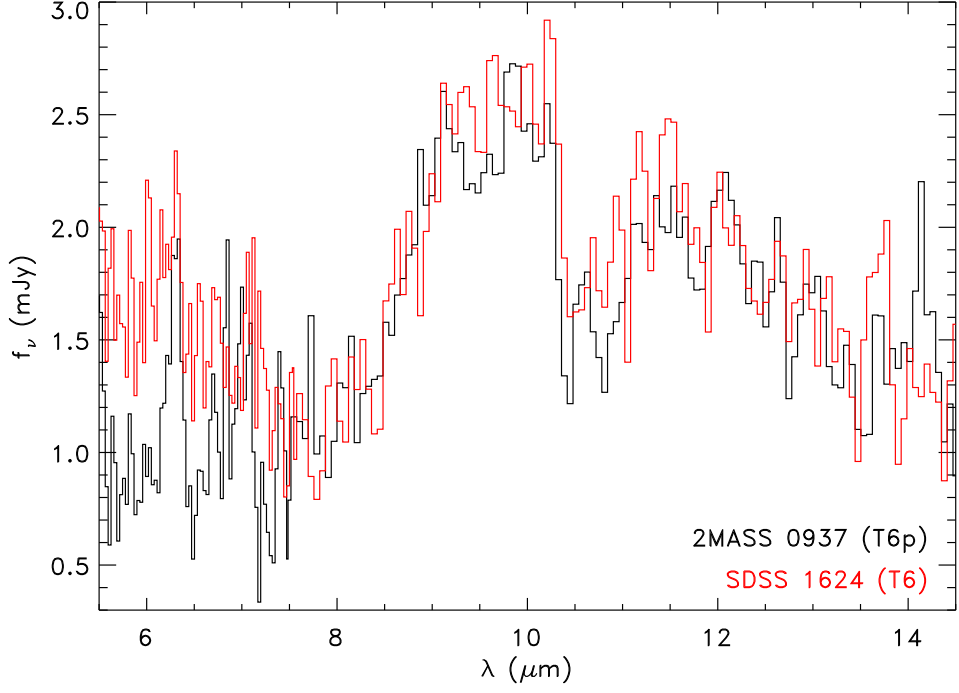}
\caption{\label{T6Comp}\textit{Top:} The IRS spectra of 2MASS
  J0937$+$2931 (T6P; \textit{black}), and 2MASS SDSS J1264$+$0029 (T6;
  \textit{red}).  The spectrum of SDSS J1624$+$0029 has been scaled by
  the ratio of the distances of the two objects to adjust its flux to
  the level that which would be observed if it were at the distance of
  2MASS J0937$+$2931 .  The spectrum of SDSS 1624 J1624$+$0029 differs
  significantly from that of 2MASS J0937$+$2931 at $\lambda \lesssim $ 7.5
  $\mu$m.}
\end{figure}

\clearpage

\begin{figure}
\includegraphics[width=5in,angle=180]{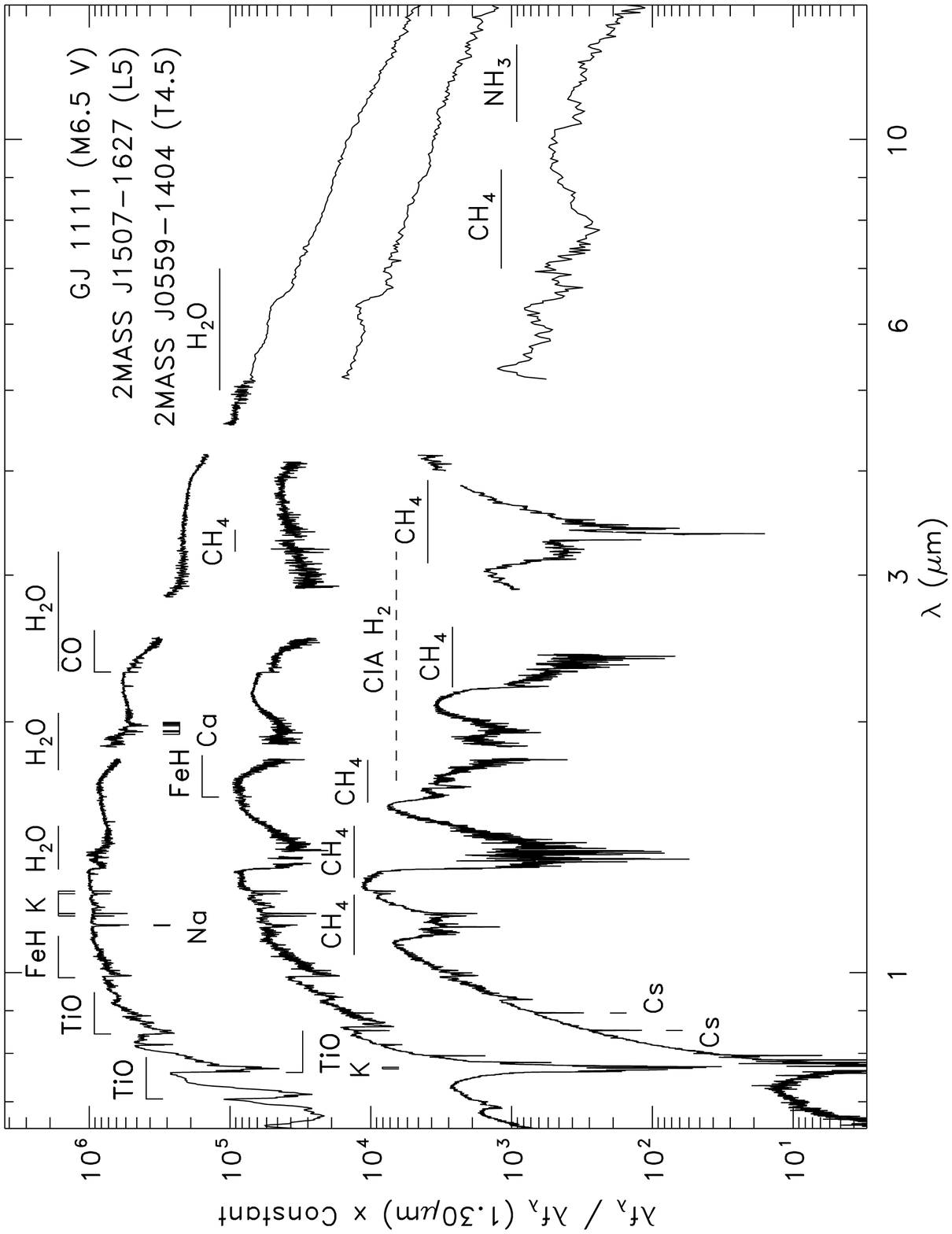}
\caption{\label{SEDSeq}The 0.65$-$14.5 spectra of GJ 1111 (M6.5 V),
  2MASS J1507$-$1627 (L5), and 2MASS J0559$-$1404 (T4.5).  The
  red-optical spectra are from \citet{1991ApJS...77..417K},
  \citet{2000AJ....119..369R}, and \citet{2003ApJ...594..510B} and the
  near-infrared spectra are from \citet{2005ApJ...623.1115C} and J.T.
  Rayner et al. (in preparation).  Note the flux density units are
  $\lambda f_\lambda$.  The spectra have been normalized to unity at
  1.3 $\mu$m and multiplied by constants.  The CIA H$_2$ absorption is
  indicated as a dashed line because it shows no distinct spectral
  features but rather a broad, smooth absorption.}
\end{figure}

\clearpage

\LongTables

\input{tab1.tex}

\clearpage

\input{tab2.tex}

\end{document}

%% file: tab1.tex
\begin{deluxetable}{lllccccc}
\tablecolumns{14}
\tabletypesize{\scriptsize} 
\tablewidth{0pc}
\tablecaption{\label{obslog}Log of the IRS Observations}
\tablehead{
\colhead{Object} & 
\colhead{Optical\tablenotemark{a}} & 
\colhead{Infrared\tablenotemark{a}} & 
\colhead{AOR Key} & 
\multicolumn{4}{c}{{Exposure Time (sec)}} \\
\cline{5-8}
\colhead{} & 
\colhead{Sp. Type} &
\colhead{Sp. Type} &
\colhead{} & 
\colhead{SL2\tablenotemark{b}} & 
\colhead{SL1} & 
\colhead{LL2} & 
\colhead{LL1}}

\startdata

Gl 229A                       & M1 V   & $\cdots$   & \phn4185856                   & \phn\phn12  & \phn\phn12 & \phn\phn30 & \phn\phn30 \\
Gl 1                          & M1.5 V & $\cdots$   & \phn3873792                   & \phn\phn24  & \phn\phn24 & $\cdots$   & $\cdots$   \\
%Gl 411                        & M2 V   & $\cdots$   & \phn3874048
G 196$-$3A                    & M2.5 V & $\cdots$   & \phn3879168                   & \phn\phn30  & \phn120    & 1220       & \phn120    \\
Gl 674                        & M2.5 V & $\cdots$   & \phn3874304                   & \phn\phn24  & \phn\phn24 & $\cdots$   & $\cdots$   \\
Gl 752A                       & M2.5 V & $\cdots$   & \phn3876864\tablenotemark{c}  & \phn\phn30  & \phn\phn30 & \phn488    & \phn488    \\
Gl 687                        & M3 V   & $\cdots$   & \phn3874560                   & \phn\phn24  & \phn\phn24 & $\cdots$   & $\cdots$   \\
Gl 849                        & M3.5 V & $\cdots$   & \phn3873024                   & \phn\phn24  & \phn\phn24 & $\cdots$   & $\cdots$   \\
GJ 1001A                      & M3.5 V & $\cdots$   & \phn4190464\tablenotemark{d}  & \phn484     & \phn484    & \phn488    & \phn976    \\
%Gl 699                        & M4 V   & $\cdots$   & 
Gl 866ABC                     & M5 V   & $\cdots$   & \phn3878912                   & \phn\phn24  & \phn\phn24 & \phn\phn24 & \phn\phn24 \\
Gl 65AB                       & M5.5 V & $\cdots$   & \phn3878400                   & \phn\phn24  & \phn\phn24 & \phn\phn24 & \phn\phn24 \\
GJ 1111                       & M6.5 V & $\cdots$   & \phn3876096                   & \phn\phn24  & \phn\phn24 & $\cdots$   & $\cdots$   \\
LHS 3003                      & M7 V   & $\cdots$   & \phn3876608                   & \phn\phn30  & \phn\phn30 &  \phn732   &  \phn732   \\ 
vB 10                         & M8 V   & $\cdots$   & 12486401                      & \phn\phn60  & \phn\phn60 & \phn488    & \phn488 \\
BRI 0021$-$0214               & M9.5 V & $\cdots$   & \phn3877632                   & \phn122     & \phn244    & $\cdots$   & $\cdots$   \\
%LSR 1610$-$0040               & M      & M6         &  12496640                     & \phn976     & \phn976    & $\cdots$   & $\cdots$   \\
2MASS J07464256$+$2000321AB   & L0.5   & L1         & \phn4186624                   & \phn968     & \phn968    & \phn244    & \phn244    \\
2MASS J14392836$+$1929149     & L1     & L1         & \phn4187136                   & \phn976     & \phn976    & $\cdots$   & $\cdots$   \\
2MASS J11083081$+$6830169     & L1     & $\cdots$   & \phn4187648                   & \phn968     & \phn968    & $\cdots$   & $\cdots$    \\
2MASS J16580380$+$7027015     & L1     & $\cdots$   & \phn4193024                   & \phn968     & \phn968    & $\cdots$   & $\cdots$    \\
Kelu$-$1AB                    & L2     & L3$\pm$1   & \phn4187904                   & \phn976     & \phn976    & $\cdots$   & $\cdots$   \\
%G196$-$3B                     & L2     & $\cdots$   & \phn4193536                   & not in \# total & need special attn. & & \\
2MASS J15065441$+$1321060     & L3     & $\cdots$   & 12496384                      & \phn976     & \phn976    & $\cdots$   & $\cdots$   \\
2MASS J16154416$+$3559005     & L3     & $\cdots$   & \phn4194048                   & \phn968     & \phn968    & \phn976    & \phn976 \\
2MASS J00361617$+$1821104     & L3.5   & L4$\pm$1   & \phn4188672                   & \phn484     & \phn484    & \phn488    & \phn976 \\
2MASS J22244381$-$0158521     & L4.5   & L3.5       & \phn4189440                   & \phn976     & \phn976    & $\cdots$   & $\cdots$   \\
2MASS J15074769$-$1627386     & L5     & L5.5       & \phn4190208                   & \phn976     & \phn976    & $\cdots$   & $\cdots$   \\
SDSS  J053951.99$-$005902.0   & L5     & L5         & \phn4190720                   & \phn968     & \phn968    & $\cdots$   & $\cdots$   \\
2MASS J12392727$+$5515371     & L5     & $\cdots$   & \phn4194304                   & \phn968     & \phn968    & $\cdots$   & $\cdots$   \\
2MASS J15150083$+$4847416     & L6.5   & $\cdots$   & \phn4190720                   & \phn976     & \phn976    & $\cdots$   & $\cdots$   \\
%2MASS J09201223$+$3517429     & L6.5   & $\cdots$   & \phn4190976                   & \phn968     & \phn968    & $\cdots$   & $\cdots$   \\
2MASS J17281150$+$3948593     & L7     & $\cdots$   &  \phn4191744                  & \phn968     & \phn968    & $\cdots$   & $\cdots$   \\
2MASS J15261405$+$2043414     & L7     & $\cdots$   & \phn4191488                   & \phn968     & \phn968    & $\cdots$   & $\cdots$   \\
2MASS J08251968$+$2115521     & L7.5   & L6         & \phn4191232                   & \phn968     & \phn968    & $\cdots$   & $\cdots$    \\ 
DENIS-P J025503.3$-$470049.0  & L8     & L9         &  \phn4192000                  & \phn484     & \phn484    & \phn488    & \phn976  \\  
%Gl 584C                       & L8     & $\cdots$   &  \phn4194560                   &  SL2  & \phn968 not included \\
%                              &        &            &                                &  SL1  & \phn968 not included \\
Gl584C                        & L8       & $\cdots$ &  \phn1249776                  & 3904        & 3904       & $\cdots$   & $\cdots$   \\
2MASS J09293364$+$3429527     & L8       & $\cdots$ &  \phn4195072                  & \phn968     & \phn968    & $\cdots$   & $\cdots$   \\
Gl 337CD                      & L8       & T0       &  12494080                     & 2928        & 2928       & $\cdots$   & $\cdots$   \\
SDSS J042348.57$-$041403.5AB  & L7.5     & T0       &  12495360                     & \phn976     & \phn976    & $\cdots$   & $\cdots$ \\
%&  \phn4195584                  & \phn488 Not in Final & \phn968 Not in Final & $\cdots$   & $\cdots$   \\
%                              &          &          
%SDSSp J083717.22$-$000018.3   & $\cdots$ & T1       &  \phn4183552                   &  SL2  & \\
%                              &          &          &                                &  SL1  & \\
%                              &          &          &                                &  LL2  &     \\
%                              &          &          &                                &  LL1  &     \\
SDSS  J125453.90$-$012247.4   & T2       & T2       &  \phn4185088                  & \phn968     & \phn968    & $\cdots$   & $\cdots$ \\
$\epsilon$ Ind Ba/Bb          & $\cdots$ & T2.5     & \phn6313730                   & \phn488     & \phn488    & \phn488    & \phn976 \\
SDSS J102109.69$-$030420.1AB  & $\cdots$ & T3       & \phn4183808                   & 3904        & 3904       & $\cdots$   & $\cdots$   \\
2MASS J05591914-1404488       & T5       & T4.5     &  \phn4183296                  & \phn968     & \phn968    & $\cdots$   & $\cdots$   \\
                         &          &          & \phn4183040                   & \phn976     & \phn976    & $\cdots$   & $\cdots$   \\
2MASS J15031961$+$2525196     & T6       & T5.5     &  \phn6314496                  & \phn976     & \phn976    & $\cdots$   & $\cdots$   \\
SDSS  J162414.37$+$002915.6   & $\cdots$ & T6       &  \phn4185600                  & 3904        & 3904       & $\cdots$   & $\cdots$   \\
2MASS J12255432$-$2739466AB   & T6       & T6       &  12496128                     & \phn976     & \phn976    & $\cdots$   & $\cdots$   \\
2MASS J09373487$+$2931409     & T7p      & T6p      &  \phn4195328                  & \phn968     & \phn968    & $\cdots$   & $\cdots$   \\
2MASS J12373919$+$6526148     & T7       & T6.5     & \phn4184832                   & \phn968     & \phn968    & $\cdots$   & $\cdots$   \\
2MASS J12171110$-$0311131     & T7       & T7.5     & \phn4184320                   & \phn976     & \phn976    & $\cdots$   & $\cdots$   \\
Gl 570D                       & T7       & T7.5     & \phn4186368                   & 3904        & 3904       & $\cdots$   & $\cdots$   \\
%2MASS 0532

\enddata

\tablenotetext{a}{Spectral types of the M dwarfs are from
  \citet{1991ApJS...77..417K}, \citet{1994AJ....108.1437H},
  \citet{1995AJ....109..797K}, \citet{1996AJ....112.2799H},
  \citet{1998Sci...282.1309R}, and J.D. Kirkpatrick (2006, private
  communication).  Spectral types of the L dwarfs are from
  \citet{1999ApJ...519..802K}, \citet{2000AJ....119..928F},
  \citet{2000AJ....119..369R}, \citet{2000AJ....120.1085G},
  \citet{2000AJ....120..447K}, \citet{2001AJ....121.3235K},
  \citet{2001AJ....122.1989W}, \citet{2003AJ....126.2421C}, and
  \citet{2006ApJ...637.1067B}.  Spectral types of the T dwarfs are from
  \citet{2003ApJ...594..510B}, and \citet{2006ApJ...637.1067B}, except
  for $\epsilon$ Ind Ba/Bb which is from \citet{2003A&A...398L..29S}.
  Spectral types for binaries are derived from unresolved spectra.
  Errors on spectral types are $\pm$0.5 subclass unless otherwise
  noted.}

\tablenotetext{b}{SL2=Short-Low Order 2, SL1=Short-Low Order1,
  LL2=Long-Low Order 2, LL1=Long-Low Order 1} \tablenotetext{c}{Original
  target was Gl 752B.}  

\tablenotetext{d}{Original target was GJ 1001B.}

\end{deluxetable}

%% file: tab2.tex
\begin{deluxetable}{lllr@{ $\pm$ }lr@{ $\pm$ }lr@{ $\pm$ }l}
\tablecolumns{14}
\tabletypesize{\scriptsize} 
\tablewidth{0pc}
\tablecaption{\label{Lboltab}Bolometric Magnitudes}
\tablehead{
\colhead{Object} & 
\colhead{Optical\tablenotemark{a}} & 
\colhead{Infrared\tablenotemark{a}} &
\multicolumn{2}{c}{{$\pi$\tablenotemark{b}}} &
\multicolumn{4}{c}{{$M_{\mathrm{bol}}$\tablenotemark{c}}} \\
\cline{6-9}
\colhead{} & 
\colhead{Sp. Type} &
\colhead{Sp. Type} & 
\multicolumn{2}{c}{{(mas)}} & 
\multicolumn{2}{c}{{Golimowski et al.}} & 
\multicolumn{2}{c}{{this work}}}

\startdata

Gl 229A                       & M1 V   & $\cdots$   & 173.17 & 1.10 & 7.97  & 0.09 & \phn7.92 & 0.04 \\ 
BRI 0021$-$0214               & M9.5 V & $\cdots$   & 84.2   & 2.6  & 13.37 & 0.10 & 13.45 & 0.08 \\
2MASS J07464256$+$2000321AB   & L0.5   & L1         & 81.9   & 0.3  & 13.26 & 0.07 & 13.29 & 0.04 \\
2MASS J14392836$+$1929149     & L1     & L1         & 69.6   & 0.5  & 13.88 & 0.07 & 14.01 & 0.04 \\
Kelu$-$1AB                    & L2     & L3$\pm$1   & 53.6   & 2.0  & 13.74 & 0.11 & 13.78 & 0.11 \\
2MASS J00361617$+$1821104     & L3.5   & L4$\pm$1   & 114.2  & 0.8  & 14.67 & 0.07 & 14.64 & 0.04 \\
2MASS J22244381$-$0158521     & L4.5   & L3.5$\pm$1 & 87.02  & 0.89 & 15.14 & 0.07 & 15.16 & 0.04 \\
2MASS J1507476$-$162738       & L5     & L5.5       & 136.4  & 0.6  & 15.16 & 0.07 & 15.28 & 0.04 \\
SDSS  J053951.99$-$005902.0   & L5     & L5         & 76.12  & 2.17 & 15.12 & 0.09 & 15.21 & 0.07 \\
2MASS J08251968$+$2115521     & L7.5   & L6         & 94.22  & 0.99 & 16.10 & 0.07 & 16.13 & 0.04 \\
%SDSSp J042348.57$-$041403.5AB & L7.5   & T0         & 65.93  & 1.70 & 15.11 & 0.10 & \\
SDSS  J125453.90$-$012247.4   & T2     & T2         & 73.96  & 1.59 & 16.08 & 0.10 & 16.22 & 0.06 \\
2MASS J05591914-1404488       & T5     & T4.5       & 96.73  & 0.96 & 16.07 & 0.13 & 16.17 & 0.04 \\

\enddata
\tablenotetext{a}{Spectral types of the M dwarfs are from
  \citet{1991ApJS...77..417K}, \citet{1994AJ....108.1437H},
  \citet{1995AJ....109..797K}, \citet{1996AJ....112.2799H},
  \citet{1998Sci...282.1309R}, and J.D. Kirkpatrick (2006, private
  communication).  Spectral types of the L dwarfs are from
  \citet{1999ApJ...519..802K}, \citet{2000AJ....119..928F},
  \citet{2000AJ....119..369R}, \citet{2000AJ....120.1085G},
  \citet{2000AJ....120..447K}, \citet{2001AJ....121.3235K},
  \citet{2001AJ....122.1989W}, \citet{2003AJ....126.2421C}, and
  \citet{2006ApJ...637.1067B}.  The spectral types of the T dwarfs are
  from \citet{2003ApJ...594..510B}, and \citet{2006ApJ...637.1067B},
  except for the spectral type of $\epsilon$ Ind Ba/Bb which is from
  \citet{2003A&A...398L..29S}.  Spectral types for binaries are derived
  from unresolved spectra.  Errors on spectral types are $\pm$0.5
  subclass unless otherwise noted.}

\tablenotetext{b}{The trigonometric parallaxes are from
  \citet{2004AJ....127.3516G} and were taken from
  \citet{1995gcts.book.....V}, \citet{1997A&A...323L..49P},
  \citet{1995AJ....110.3014T}, \citet{2002AJ....124.1170D},
  \citet{2004AJ....127.2948V}, and \cite{2003AJ....126..975T}.}

\tablenotetext{c}{$M_{\mathrm{bol}}$ = $-$2.5 $\log f_{\mathrm{bol}}$ + 5$\log \pi$ $-$13.978 assuming  $L_{\odot} = 3.86\times10^{26}$ W and $M_{\mathrm{bol}\odot} = +4.75$.}

\end{deluxetable}

%% file: ms.bbl
\begin{thebibliography}{79}
\expandafter\ifx\csname natexlab\endcsname\relax\def\natexlab#1{#1}\fi

\bibitem[{{Ackerman} \& {Marley}(2001)}]{2001ApJ...556..872A}
{Ackerman}, A.~S. \& {Marley}, M.~S. 2001, \apj, 556, 872

\bibitem[{{Allard} {et~al.}(2001){Allard}, {Hauschildt}, {Alexander},
  {Tamanai}, \& {Schweitzer}}]{2001ApJ...556..357A}
{Allard}, F., {Hauschildt}, P.~H., {Alexander}, D.~R., {Tamanai}, A., \&
  {Schweitzer}, A. 2001, \apj, 556, 357

\bibitem[{{Basri}(2000)}]{2000ARA&A..38..485B}
{Basri}, G. 2000, \araa, 38, 485

\bibitem[{{Basri} {et~al.}(1996){Basri}, {Marcy}, \&
  {Graham}}]{1996ApJ...458..600B}
{Basri}, G., {Marcy}, G.~W., \& {Graham}, J.~R. 1996, \apj, 458, 600

\bibitem[{{Borysow}(2002)}]{2002A&A...390..779B}
{Borysow}, A. 2002, \aap, 390, 779

\bibitem[{{Borysow} \& {Frommhold}(1986)}]{1986ApJ...303..495B}
{Borysow}, A. \& {Frommhold}, L. 1986, \apj, 303, 495

\bibitem[{{Burgasser}(2001)}]{2001PhDT.......116B}
{Burgasser}, A.~J. 2001, Ph.D.~Thesis,~Caltech

\bibitem[{{Burgasser} {et~al.}(2005{\natexlab{a}}){Burgasser}, {Burrows}, \&
  D.}]{burgasser05b}
{Burgasser}, A.~J., {Burrows}, A., \& D., K.~J. 2005{\natexlab{a}}, astro-ph
  0510707

\bibitem[{{Burgasser} {et~al.}(2006){Burgasser}, {Geballe}, {Leggett},
  {Kirkpatrick}, \& {Golimowski}}]{2006ApJ...637.1067B}
{Burgasser}, A.~J., {Geballe}, T.~R., {Leggett}, S.~K., {Kirkpatrick}, J.~D.,
  \& {Golimowski}, D.~A. 2006, \apj, 637, 1067

\bibitem[{{Burgasser} {et~al.}(2002){Burgasser}, {Kirkpatrick}, {Brown},
  {Reid}, {Burrows}, {Liebert}, {Matthews}, {Gizis}, {Dahn}, {Monet}, {Cutri},
  \& {Skrutskie}}]{2002ApJ...564..421B}
{Burgasser}, A.~J., {Kirkpatrick}, J.~D., {Brown}, M.~E., {Reid}, I.~N.,
  {Burrows}, A., {Liebert}, J., {Matthews}, K., {Gizis}, J.~E., {Dahn}, C.~C.,
  {Monet}, D.~G., {Cutri}, R.~M., \& {Skrutskie}, M.~F. 2002, \apj, 564, 421

\bibitem[{{Burgasser} {et~al.}(2003){Burgasser}, {Kirkpatrick}, {Liebert}, \&
  {Burrows}}]{2003ApJ...594..510B}
{Burgasser}, A.~J., {Kirkpatrick}, J.~D., {Liebert}, J., \& {Burrows}, A. 2003,
  \apj, 594, 510

\bibitem[{{Burgasser} {et~al.}(2005{\natexlab{b}}){Burgasser}, {Kirkpatrick},
  \& {Lowrance}}]{2005AJ....129.2849B}
{Burgasser}, A.~J., {Kirkpatrick}, J.~D., \& {Lowrance}, P.~J.
  2005{\natexlab{b}}, \aj, 129, 2849

\bibitem[{{Burgasser} {et~al.}(2005{\natexlab{c}}){Burgasser}, {Reid},
  {Leggett}, {Kirkpatrick}, {Liebert}, \& {Burrows}}]{2005ApJ...634L.177B}
{Burgasser}, A.~J., {Reid}, I.~N., {Leggett}, S.~K., {Kirkpatrick}, J.~D.,
  {Liebert}, J., \& {Burrows}, A. 2005{\natexlab{c}}, \apjl, 634, L177

\bibitem[{{Burrows} {et~al.}(2001){Burrows}, {Hubbard}, {Lunine}, \&
  {Liebert}}]{2001RvMP...73..719B}
{Burrows}, A., {Hubbard}, W.~B., {Lunine}, J.~I., \& {Liebert}, J. 2001,
  Reviews of Modern Physics, 73, 719

\bibitem[{{Burrows} \& {Sharp}(1999)}]{1999ApJ...512..843B}
{Burrows}, A. \& {Sharp}, C.~M. 1999, \apj, 512, 843

\bibitem[{{Burrows} {et~al.}(2006){Burrows}, {Sudarsky}, \&
  {Hubeny}}]{2006ApJ...640.1063B}
{Burrows}, A., {Sudarsky}, D., \& {Hubeny}, I. 2006, \apj, 640, 1063

\bibitem[{{Chabrier} \& {Baraffe}(2000)}]{2000ARA&A..38..337C}
{Chabrier}, G. \& {Baraffe}, I. 2000, \araa, 38, 337

\bibitem[{{Cohen} {et~al.}(2003){Cohen}, {Megeath}, {Hammersley},
  {Mart{\'{\i}}n-Luis}, \& {Stauffer}}]{2003AJ....125.2645C}
{Cohen}, M., {Megeath}, S.~T., {Hammersley}, P.~L., {Mart{\'{\i}}n-Luis}, F.,
  \& {Stauffer}, J. 2003, \aj, 125, 2645

\bibitem[{{Creech-Eakman} {et~al.}(2004){Creech-Eakman}, {Orton}, {Serabyn}, \&
  {Hayward}}]{2004ApJ...602L.129C}
{Creech-Eakman}, M.~J., {Orton}, G.~S., {Serabyn}, E., \& {Hayward}, T.~L.
  2004, \apjl, 602, L129

\bibitem[{{Cruz} {et~al.}(2003){Cruz}, {Reid}, {Liebert}, {Kirkpatrick}, \&
  {Lowrance}}]{2003AJ....126.2421C}
{Cruz}, K.~L., {Reid}, I.~N., {Liebert}, J., {Kirkpatrick}, J.~D., \&
  {Lowrance}, P.~J. 2003, \aj, 126, 2421

\bibitem[{{Cushing} {et~al.}(2005){Cushing}, {Rayner}, \&
  {Vacca}}]{2005ApJ...623.1115C}
{Cushing}, M.~C., {Rayner}, J.~T., \& {Vacca}, W.~D. 2005, \apj, 623, 1115

\bibitem[{{Cushing} {et~al.}(2004){Cushing}, {Vacca}, \&
  {Rayner}}]{2004PASP..116..362C}
{Cushing}, M.~C., {Vacca}, W.~D., \& {Rayner}, J.~T. 2004, \pasp, 116, 362

\bibitem[{{Dahn} {et~al.}(2002){Dahn}, {Harris}, {Vrba}, {Guetter}, {Canzian},
  {Henden}, {Levine}, {Luginbuhl}, {Monet}, {Monet}, {Pier}, {Stone}, {Walker},
  {Burgasser}, {Gizis}, {Kirkpatrick}, {Liebert}, \&
  {Reid}}]{2002AJ....124.1170D}
{Dahn}, C.~C., et al. 2002, \aj, 124, 1170

\bibitem[{{Epchtein} {et~al.}(1997){Epchtein}, {de Batz}, {Capoani},
  {Chevallier}, {Copet}, {Fouque}, {Lacombe}, {Le Bertre}, {Pau}, {Rouan},
  {Ruphy}, {Simon}, {Tiphene}, {Burton}, {Bertin}, {Deul}, {Habing},
  {Borsenberger}, {Dennefeld}, {Guglielmo}, {Loup}, {Mamon}, {Ng}, {Omont},
  {Provost}, {Renault}, {Tanguy}, {Kimeswenger}, {Kienel}, {Garzon}, {Persi},
  {Ferrari-Toniolo}, {Robin}, {Paturel}, {Vauglin}, {Forveille}, {Delfosse},
  {Hron}, {Schultheis}, {Appenzeller}, {Wagner}, {Balazs}, {Holl}, {Lepine},
  {Boscolo}, {Picazzio}, {Duc}, \& {Mennessier}}]{1997Msngr..87...27E}
{Epchtein}, N., et al. 1997, The Messenger, 87,
  27

\bibitem[{{Fan} {et~al.}(2000){Fan}, {Knapp}, {Strauss}, {Gunn}, {Lupton},
  {Ivezi{\'c}}, {Rockosi}, {Yanny}, {Kent}, {Schneider}, {Kirkpatrick},
  {Annis}, {Bastian}, {Berman}, {Brinkmann}, {Csabai}, {Federwitz}, {Fukugita},
  {Gurbani}, {Hennessy}, {Hindsley}, {Ichikawa}, {Lamb}, {Lindenmeyer},
  {Mantsch}, {McKay}, {Munn}, {Nash}, {Okamura}, {Pauls}, {Pier},
  {Rechenmacher}, {Rivetta}, {Sergey}, {Stoughton}, {Szalay}, {Szokoly},
  {Tucker}, {York}, \& {The SDSS Collaboration}}]{2000AJ....119..928F}
{Fan}, X., et al. 2000, \aj, 119, 928

\bibitem[{{Fazio} {et~al.}(2004){Fazio}, {Hora}, {Allen}, {Ashby}, {Barmby},
  {Deutsch}, {Huang}, {Kleiner}, {Marengo}, {Megeath}, {Melnick}, {Pahre},
  {Patten}, {Polizotti}, {Smith}, {Taylor}, {Wang}, {Willner}, {Hoffmann},
  {Pipher}, {Forrest}, {McMurty}, {McCreight}, {McKelvey}, {McMurray}, {Koch},
  {Moseley}, {Arendt}, {Mentzell}, {Marx}, {Losch}, {Mayman}, {Eichhorn},
  {Krebs}, {Jhabvala}, {Gezari}, {Fixsen}, {Flores}, {Shakoorzadeh}, {Jungo},
  {Hakun}, {Workman}, {Karpati}, {Kichak}, {Whitley}, {Mann}, {Tollestrup},
  {Eisenhardt}, {Stern}, {Gorjian}, {Bhattacharya}, {Carey}, {Nelson},
  {Glaccum}, {Lacy}, {Lowrance}, {Laine}, {Reach}, {Stauffer}, {Surace},
  {Wilson}, {Wright}, {Hoffman}, {Domingo}, \& {Cohen}}]{2004ApJS..154...10F}
{Fazio}, G.~G., et al. 2004, \apjs, 154, 10

\bibitem[{{Fegley} \& {Lodders}(1996)}]{1996ApJ...472L..37F}
{Fegley}, B.~J. \& {Lodders}, K. 1996, \apjl, 472, L37

\bibitem[{{Geballe} {et~al.}(2002){Geballe}, {Knapp}, {Leggett}, {Fan},
  {Golimowski}, {Anderson}, {Brinkmann}, {Csabai}, {Gunn}, {Hawley},
  {Hennessy}, {Henry}, {Hill}, {Hindsley}, {Ivezi{\'c}}, {Lupton}, {McDaniel},
  {Munn}, {Narayanan}, {Peng}, {Pier}, {Rockosi}, {Schneider}, {Smith},
  {Strauss}, {Tsvetanov}, {Uomoto}, {York}, \& {Zheng}}]{2002ApJ...564..466G}
{Geballe}, T.~R., et al. 2002, \apj, 564, 466

\bibitem[{{Gizis} {et~al.}(2000){Gizis}, {Monet}, {Reid}, {Kirkpatrick},
  {Liebert}, \& {Williams}}]{2000AJ....120.1085G}
{Gizis}, J.~E., {Monet}, D.~G., {Reid}, I.~N., {Kirkpatrick}, J.~D., {Liebert},
  J., \& {Williams}, R.~J. 2000, \aj, 120, 1085

\bibitem[{{Golimowski} {et~al.}(2004){Golimowski}, {Leggett}, {Marley}, {Fan},
  {Geballe}, {Knapp}, {Vrba}, {Henden}, {Luginbuhl}, {Guetter}, {Munn},
  {Canzian}, {Zheng}, {Tsvetanov}, {Chiu}, {Glazebrook}, {Hoversten},
  {Schneider}, \& {Brinkmann}}]{2004AJ....127.3516G}
{Golimowski}, D.~A., et al. 2004, \aj, 127, 3516

\bibitem[{{Hanner} {et~al.}(1994){Hanner}, {Lynch}, \&
  {Russell}}]{1994ApJ...425..274H}
{Hanner}, M.~S., {Lynch}, D.~K., \& {Russell}, R.~W. 1994, \apj, 425, 274

\bibitem[{{Hawley} {et~al.}(1996){Hawley}, {Gizis}, \&
  {Reid}}]{1996AJ....112.2799H}
{Hawley}, S.~L., {Gizis}, J.~E., \& {Reid}, I.~N. 1996, \aj, 112, 2799

\bibitem[{{Helling} {et~al.}(2006){Helling}, {Thi}, {Woitke}, \&
  {Fridlund}}]{helling06}
{Helling}, C., {Thi}, W.-F., {Woitke}, P., \& {Fridlund}, M. 2006, in press
  (astro-ph 0603341)

\bibitem[{{Henry} {et~al.}(1994){Henry}, {Kirkpatrick}, \&
  {Simons}}]{1994AJ....108.1437H}
{Henry}, T.~J., {Kirkpatrick}, J.~D., \& {Simons}, D.~A. 1994, \aj, 108, 1437

\bibitem[{{Herzberg}(1945)}]{1945msms.book.....H}
{Herzberg}, G. 1945, {Molecular spectra and molecular structure. Vol.2:
  Infrared and Raman spectra of polyatomic molecules} (New York: Van Nostrand,
  Reinhold, 1945)

\bibitem[{{Higdon} {et~al.}(2004){Higdon}, {Devost}, {Higdon}, {Brandl},
  {Houck}, {Hall}, {Barry}, {Charmandaris}, {Smith}, {Sloan}, \&
  {Green}}]{2004PASP..116..975H}
{Higdon}, S.~J.~U., et al. 2004, \pasp, 116, 975

\bibitem[{{Houck} {et~al.}(2004){Houck}, {Roellig}, {van Cleve}, {Forrest},
  {Herter}, {Lawrence}, {Matthews}, {Reitsema}, {Soifer}, {Watson}, {Weedman},
  {Huisjen}, {Troeltzsch}, {Barry}, {Bernard-Salas}, {Blacken}, {Brandl},
  {Charmandaris}, {Devost}, {Gull}, {Hall}, {Henderson}, {Higdon}, {Pirger},
  {Schoenwald}, {Sloan}, {Uchida}, {Appleton}, {Armus}, {Burgdorf},
  {Fajardo-Acosta}, {Grillmair}, {Ingalls}, {Morris}, \&
  {Teplitz}}]{2004ApJS..154...18H}
{Houck}, J.~R., et al. 2004, \apjs, 154, 18

\bibitem[{{Kirkpatrick} {et~al.}(2001){Kirkpatrick}, {Dahn}, {Monet}, {Reid},
  {Gizis}, {Liebert}, \& {Burgasser}}]{2001AJ....121.3235K}
{Kirkpatrick}, J.~D., {Dahn}, C.~C., {Monet}, D.~G., {Reid}, I.~N., {Gizis},
  J.~E., {Liebert}, J., \& {Burgasser}, A.~J. 2001, \aj, 121, 3235

\bibitem[{{Kirkpatrick} {et~al.}(1991){Kirkpatrick}, {Henry}, \&
  {McCarthy}}]{1991ApJS...77..417K}
{Kirkpatrick}, J.~D., {Henry}, T.~J., \& {McCarthy}, D.~W. 1991, \apjs, 77, 417

\bibitem[{{Kirkpatrick} {et~al.}(1995){Kirkpatrick}, {Henry}, \&
  {Simons}}]{1995AJ....109..797K}
{Kirkpatrick}, J.~D., {Henry}, T.~J., \& {Simons}, D.~A. 1995, \aj, 109, 797

\bibitem[{{Kirkpatrick} {et~al.}(1999){Kirkpatrick}, {Reid}, {Liebert},
  {Cutri}, {Nelson}, {Beichman}, {Dahn}, {Monet}, {Gizis}, \&
  {Skrutskie}}]{1999ApJ...519..802K}
{Kirkpatrick}, J.~D., {Reid}, I.~N., {Liebert}, J., {Cutri}, R.~M., {Nelson},
  B., {Beichman}, C.~A., {Dahn}, C.~C., {Monet}, D.~G., {Gizis}, J.~E., \&
  {Skrutskie}, M.~F. 1999, \apj, 519, 802

\bibitem[{{Kirkpatrick} {et~al.}(2000){Kirkpatrick}, {Reid}, {Liebert},
  {Gizis}, {Burgasser}, {Monet}, {Dahn}, {Nelson}, \&
  {Williams}}]{2000AJ....120..447K}
{Kirkpatrick}, J.~D., {Reid}, I.~N., {Liebert}, J., {Gizis}, J.~E.,
  {Burgasser}, A.~J., {Monet}, D.~G., {Dahn}, C.~C., {Nelson}, B., \&
  {Williams}, R.~J. 2000, \aj, 120, 447

\bibitem[{{Knapp} {et~al.}(2004){Knapp}, {Leggett}, {Fan}, {Marley}, {Geballe},
  {Golimowski}, {Finkbeiner}, {Gunn}, {Hennawi}, {Ivezi{\'c}}, {Lupton},
  {Schlegel}, {Strauss}, {Tsvetanov}, {Chiu}, {Hoversten}, {Glazebrook},
  {Zheng}, {Hendrickson}, {Williams}, {Uomoto}, {Vrba}, {Henden}, {Luginbuhl},
  {Guetter}, {Munn}, {Canzian}, {Schneider}, \&
  {Brinkmann}}]{2004AJ....127.3553K}
{Knapp}, G.~R., et al. 2004, \aj, 127, 3553

\bibitem[{{Leggett} {et~al.}(2000){Leggett}, {Allard}, {Dahn}, {Hauschildt},
  {Kerr}, \& {Rayner}}]{2000ApJ...535..965L}
{Leggett}, S.~K., {Allard}, F., {Dahn}, C., {Hauschildt}, P.~H., {Kerr}, T.~H.,
  \& {Rayner}, J. 2000, \apj, 535, 965

\bibitem[{{Leggett} {et~al.}(2001){Leggett}, {Allard}, {Geballe}, {Hauschildt},
  \& {Schweitzer}}]{2001ApJ...548..908L}
{Leggett}, S.~K., {Allard}, F., {Geballe}, T.~R., {Hauschildt}, P.~H., \&
  {Schweitzer}, A. 2001, \apj, 548, 908

\bibitem[{{Leggett} {et~al.}(2002){Leggett}, {Golimowski}, {Fan}, {Geballe},
  {Knapp}, {Brinkmann}, {Csabai}, {Gunn}, {Hawley}, {Henry}, {Hindsley},
  {Ivezi{\'c}}, {Lupton}, {Pier}, {Schneider}, {Smith}, {Strauss}, {Uomoto}, \&
  {York}}]{2002ApJ...564..452L}
{Leggett}, S.~K., et al. 2002, \apj, 564, 452

\bibitem[{{Lodders}(1999)}]{1999ApJ...519..793L}
{Lodders}, K. 1999, \apj, 519, 793

\bibitem[{{Lodders}(2002)}]{2002ApJ...577..974L}
---. 2002, \apj, 577, 974

\bibitem[{{Lodders} \& {Fegley}(2002)}]{2002Icar..155..393L}
{Lodders}, K. \& {Fegley}, B. 2002, Icarus, 155, 393

\bibitem[{{Marley}(2000)}]{2000fgpc.conf..152M}
{Marley}, M. 2000, in ASP Conf. Ser. 212: From Giant Planets to Cool Stars, 152

\bibitem[{{Marley} {et~al.}(2002){Marley}, {Seager}, {Saumon}, {Lodders},
  {Ackerman}, {Freedman}, \& {Fan}}]{2002ApJ...568..335M}
{Marley}, M.~S., {Seager}, S., {Saumon}, D., {Lodders}, K., {Ackerman}, A.~S.,
  {Freedman}, R.~S., \& {Fan}, X. 2002, \apj, 568, 335

\bibitem[{{Matthews} {et~al.}(1996){Matthews}, {Nakajima}, {Kulkarni}, \&
  {Oppenheimer}}]{1996AJ....112.1678M}
{Matthews}, K., {Nakajima}, T., {Kulkarni}, S.~R., \& {Oppenheimer}, B.~R.
  1996, \aj, 112, 1678

\bibitem[{{McLean} {et~al.}(2003){McLean}, {McGovern}, {Burgasser},
  {Kirkpatrick}, {Prato}, \& {Kim}}]{2003ApJ...596..561M}
{McLean}, I.~S., {McGovern}, M.~R., {Burgasser}, A.~J., {Kirkpatrick}, J.~D.,
  {Prato}, L., \& {Kim}, S.~S. 2003, \apj, 596, 561

\bibitem[{{Nakajima} {et~al.}(1995){Nakajima}, {Oppenheimer}, {Kulkarni},
  {Golimowski}, {Matthews}, \& {Durrance}}]{1995Natur.378..463N}
{Nakajima}, T., {Oppenheimer}, B.~R., {Kulkarni}, S.~R., {Golimowski}, D.~A.,
  {Matthews}, K., \& {Durrance}, S.~T. 1995, \nat, 378, 463

\bibitem[{{Noll} {et~al.}(2000){Noll}, {Geballe}, {Leggett}, \&
  {Marley}}]{2000ApJ...541L..75N}
{Noll}, K.~S., {Geballe}, T.~R., {Leggett}, S.~K., \& {Marley}, M.~S. 2000,
  \apjl, 541, L75

\bibitem[{{Noll} {et~al.}(1997){Noll}, {Geballe}, \&
  {Marley}}]{1997ApJ...489L..87N}
{Noll}, K.~S., {Geballe}, T.~R., \& {Marley}, M.~S. 1997, \apjl, 489, L87

\bibitem[{{Oppenheimer} {et~al.}(1998){Oppenheimer}, {Kulkarni}, {Matthews}, \&
  {van Kerkwijk}}]{1998ApJ...502..932O}
{Oppenheimer}, B.~R., {Kulkarni}, S.~R., {Matthews}, K., \& {van Kerkwijk},
  M.~H. 1998, \apj, 502, 932

\bibitem[{{Perryman} {et~al.}(1997){Perryman}, {Lindegren}, {Kovalevsky},
  {Hoeg}, {Bastian}, {Bernacca}, {Cr{\'e}z{\'e}}, {Donati}, {Grenon}, {van
  Leeuwen}, {van der Marel}, {Mignard}, {Murray}, {Le Poole}, {Schrijver},
  {Turon}, {Arenou}, {Froeschl{\'e}}, \& {Petersen}}]{1997A&A...323L..49P}
{Perryman}, M.~A.~C., et al. 1997, \aap, 323, L49

\bibitem[{{Reach} {et~al.}(2005){Reach}, {Megeath}, {Cohen}, {Hora}, {Carey},
  {Surace}, {Willner}, {Barmby}, {Wilson}, {Glaccum}, {Lowrance}, {Marengo}, \&
  {Fazio}}]{2005PASP..117..978R}
{Reach}, W.~T., {Megeath}, S.~T., {Cohen}, M., {Hora}, J., {Carey}, S.,
  {Surace}, J., {Willner}, S.~P., {Barmby}, P., {Wilson}, G., {Glaccum}, W.,
  {Lowrance}, P., {Marengo}, M., \& {Fazio}, G.~G. 2005, \pasp, 117, 978

\bibitem[{{Rebolo} {et~al.}(1996){Rebolo}, {Martin}, {Basri}, {Marcy}, \&
  {Zapatero-Osorio}}]{1996ApJ...469L..53R}
{Rebolo}, R., {Martin}, E.~L., {Basri}, G., {Marcy}, G.~W., \&
  {Zapatero-Osorio}, M.~R. 1996, \apjl, 469, L53

\bibitem[{{Rebolo} {et~al.}(1998){Rebolo}, {Zapatero Osorio}, {Madruga},
  {Bejar}, {Arribas}, \& {Licandro}}]{1998Sci...282.1309R}
{Rebolo}, R., {Zapatero Osorio}, M.~R., {Madruga}, S., {Bejar}, V.~J.~S.,
  {Arribas}, S., \& {Licandro}, J. 1998, Science, 282, 1309

\bibitem[{{Reid} {et~al.}(2000){Reid}, {Kirkpatrick}, {Gizis}, {Dahn}, {Monet},
  {Williams}, {Liebert}, \& {Burgasser}}]{2000AJ....119..369R}
{Reid}, I.~N., {Kirkpatrick}, J.~D., {Gizis}, J.~E., {Dahn}, C.~C., {Monet},
  D.~G., {Williams}, R.~J., {Liebert}, J., \& {Burgasser}, A.~J. 2000, \aj,
  119, 369

\bibitem[{{Roellig} {et~al.}(2004){Roellig}, {Van Cleve}, {Sloan}, {Wilson},
  {Saumon}, {Leggett}, {Marley}, {Cushing}, {Kirkpatrick}, {Mainzer}, \&
  {Houck}}]{2004ApJS..154..418R}
{Roellig}, T.~L., et al. 2004, \apjs, 154,
  418

\bibitem[{{Saumon} {et~al.}(2003{\natexlab{a}}){Saumon}, {Marley}, \&
  {Lodders}}]{saumon03}
{Saumon}, D., {Marley}, M.~S., \& {Lodders}, K. 2003{\natexlab{a}}, astro-ph
  0310805

\bibitem[{{Saumon} {et~al.}(2003{\natexlab{b}}){Saumon}, {Marley}, {Lodders},
  \& {Freedman}}]{2003IAUS..211..345S}
{Saumon}, D., {Marley}, M.~S., {Lodders}, K., \& {Freedman}, R.~S.
  2003{\natexlab{b}}, in IAU Symp.~211, ~Brown Dwarfs,~ed. E. Mart{\'{\i}}n
  (San Francisco:~ASP), 345

\bibitem[{{Scholz} {et~al.}(2003){Scholz}, {McCaughrean}, {Lodieu}, \&
  {Kuhlbrodt}}]{2003A&A...398L..29S}
{Scholz}, R.-D., {McCaughrean}, M.~J., {Lodieu}, N., \& {Kuhlbrodt}, B. 2003,
  \aap, 398, L29

\bibitem[{{Skrutskie} {et~al.}(2006){Skrutskie}, {Cutri}, {Stiening},
  {Weinberg}, {Schneider}, {Carpenter}, {Beichman}, {Capps}, {Chester},
  {Elias}, {Huchra}, {Liebert}, {Lonsdale}, {Monet}, {Price}, {Seitzer},
  {Jarrett}, {Kirkpatrick}, {Gizis}, {Howard}, {Evans}, {Fowler}, {Fullmer},
  {Hurt}, {Light}, {Kopan}, {Marsh}, {McCallon}, {Tam}, {Van Dyk}, \&
  {Wheelock}}]{2006AJ....131.1163S}
{Skrutskie}, M.~F., et al. 2006, \aj, 131, 1163

\bibitem[{{Stephens} {et~al.}(2001){Stephens}, {Marley}, {Noll}, \&
  {Chanover}}]{2001ApJ...556L..97S}
{Stephens}, D.~C., {Marley}, M.~S., {Noll}, K.~S., \& {Chanover}, N. 2001,
  \apjl, 556, L97

\bibitem[{{Sterzik} {et~al.}(2005){Sterzik}, {Pantin}, {Hartung}, {Huelamo},
  {K{\"a}ufl}, {Kaufer}, {Melo}, {N{\"u}rnberger}, {Siebenmorgen}, \&
  {Smette}}]{2005A&A...436L..39S}
{Sterzik}, M.~F., {Pantin}, E., {Hartung}, M., {Huelamo}, N., {K{\"a}ufl},
  H.~U., {Kaufer}, A., {Melo}, C., {N{\"u}rnberger}, D., {Siebenmorgen}, R., \&
  {Smette}, A. 2005, \aap, 436, L39

\bibitem[{{Tinney} {et~al.}(2003){Tinney}, {Burgasser}, \&
  {Kirkpatrick}}]{2003AJ....126..975T}
{Tinney}, C.~G., {Burgasser}, A.~J., \& {Kirkpatrick}, J.~D. 2003, \aj, 126,
  975

\bibitem[{{Tinney} {et~al.}(1995){Tinney}, {Reid}, {Gizis}, \&
  {Mould}}]{1995AJ....110.3014T}
{Tinney}, C.~G., {Reid}, I.~N., {Gizis}, J., \& {Mould}, J.~R. 1995, \aj, 110,
  3014

\bibitem[{{Toppani} {et~al.}(2004){Toppani}, {Libourel}, {Robert}, {Ghanbaja},
  \& {Zimmermann}}]{2004LPI....35.1726T}
{Toppani}, A., {Libourel}, G., {Robert}, F., {Ghanbaja}, J., \& {Zimmermann},
  L. 2004, in Lunar and Planetary Institute Conference Abstracts, 1726

\bibitem[{{van Altena} {et~al.}(1995){van Altena}, {Lee}, \&
  {Hoffleit}}]{1995gcts.book.....V}
{van Altena}, W.~F., {Lee}, J.~T., \& {Hoffleit}, E.~D. 1995, {The Yale
  Parallax Catalog} (4th ed.; New Haven: Yale Univ. Observatory)

\bibitem[{{Vrba} {et~al.}(2004){Vrba}, {Henden}, {Luginbuhl}, {Guetter},
  {Munn}, {Canzian}, {Burgasser}, {Kirkpatrick}, {Fan}, {Geballe},
  {Golimowski}, {Knapp}, {Leggett}, {Schneider}, \&
  {Brinkmann}}]{2004AJ....127.2948V}
{Vrba}, F.~J., et al. 2004, \aj, 127, 2948

\bibitem[{{Weck} {et~al.}(2004){Weck}, {Schweitzer}, {Kirby}, {Hauschildt}, \&
  {Stancil}}]{2004ApJ...613..567W}
{Weck}, P.~F., {Schweitzer}, A., {Kirby}, K., {Hauschildt}, P.~H., \&
  {Stancil}, P.~C. 2004, \apj, 613, 567

\bibitem[{{Werner} {et~al.}(2004){Werner}, {Roellig}, {Low}, {Rieke}, {Rieke},
  {Hoffmann}, {Young}, {Houck}, {Brandl}, {Fazio}, {Hora}, {Gehrz}, {Helou},
  {Soifer}, {Stauffer}, {Keene}, {Eisenhardt}, {Gallagher}, {Gautier}, {Irace},
  {Lawrence}, {Simmons}, {Van Cleve}, {Jura}, {Wright}, \&
  {Cruikshank}}]{2004ApJS..154....1W}
{Werner}, M.~W., et al. 2004, \apjs, 154, 1

\bibitem[{{Wilson} {et~al.}(2001){Wilson}, {Kirkpatrick}, {Gizis}, {Skrutskie},
  {Monet}, \& {Houck}}]{2001AJ....122.1989W}
{Wilson}, J.~C., {Kirkpatrick}, J.~D., {Gizis}, J.~E., {Skrutskie}, M.~F.,
  {Monet}, D.~G., \& {Houck}, J.~R. 2001, \aj, 122, 1989

\bibitem[{{Woitke} \& {Helling}(2004)}]{2004A&A...414..335W}
{Woitke}, P. \& {Helling}, C. 2004, \aap, 414, 335

\bibitem[{{York} {et~al.}(2000){York}, {Adelman}, {Anderson}, {Anderson},
  {Annis}, {Bahcall}, {Bakken}, {Barkhouser}, {Bastian}, {Berman}, {Boroski},
  {Bracker}, {Briegel}, {Briggs}, {Brinkmann}, {Brunner}, {Burles}, {Carey},
  {Carr}, {Castander}, {Chen}, {Colestock}, {Connolly}, {Crocker}, {Csabai},
  {Czarapata}, {Davis}, {Doi}, {Dombeck}, {Eisenstein}, {Ellman}, {Elms},
  {Evans}, {Fan}, {Federwitz}, {Fiscelli}, {Friedman}, {Frieman}, {Fukugita},
  {Gillespie}, {Gunn}, {Gurbani}, {de Haas}, {Haldeman}, {Harris}, {Hayes},
  {Heckman}, {Hennessy}, {Hindsley}, {Holm}, {Holmgren}, {Huang}, {Hull},
  {Husby}, {Ichikawa}, {Ichikawa}, {Ivezi{\'c}}, {Kent}, {Kim}, {Kinney},
  {Klaene}, {Kleinman}, {Kleinman}, {Knapp}, {Korienek}, {Kron}, {Kunszt},
  {Lamb}, {Lee}, {Leger}, {Limmongkol}, {Lindenmeyer}, {Long}, {Loomis},
  {Loveday}, {Lucinio}, {Lupton}, {MacKinnon}, {Mannery}, {Mantsch}, {Margon},
  {McGehee}, {McKay}, {Meiksin}, {Merelli}, {Monet}, {Munn}, {Narayanan},
  {Nash}, {Neilsen}, {Neswold}, {Newberg}, {Nichol}, {Nicinski}, {Nonino},
  {Okada}, {Okamura}, {Ostriker}, {Owen}, {Pauls}, {Peoples}, {Peterson},
  {Petravick}, {Pier}, {Pope}, {Pordes}, {Prosapio}, {Rechenmacher}, {Quinn},
  {Richards}, {Richmond}, {Rivetta}, {Rockosi}, {Ruthmansdorfer}, {Sandford},
  {Schlegel}, {Schneider}, {Sekiguchi}, {Sergey}, {Shimasaku}, {Siegmund},
  {Smee}, {Smith}, {Snedden}, {Stone}, {Stoughton}, {Strauss}, {Stubbs},
  {SubbaRao}, {Szalay}, {Szapudi}, {Szokoly}, {Thakar}, {Tremonti}, {Tucker},
  {Uomoto}, {Vanden Berk}, {Vogeley}, {Waddell}, {Wang}, {Watanabe},
  {Weinberg}, {Yanny}, \& {Yasuda}}]{2000AJ....120.1579Y}
{York}, D.~G., et al. 2000, \aj, 120, 1579

\end{thebibliography}
